\documentstyle[preprint,aps,floats,psfig]{revtex}

\tighten

\begin{document}

\draft

\title{Perturbative expansions from Monte Carlo
simulations at weak coupling: Wilson loops and
the static-quark self-energy}

\author{H. D. Trottier,${}^1$ N. H. Shakespeare,${}^1$
G. P. Lepage,${}^2$ and P. B. Mackenzie${}^3$}

\address{\rm{(HPQCD Collaboration)}}

\address{${}^1$Physics Department, Simon Fraser University,
Burnaby, B.C., Canada V5A 1S6}

\address{${}^2$Newman Laboratory of Nuclear Studies,
Cornell University, Ithaca, NY, 14853}

\address{${}^3$Fermilab, P.O. Box 500, M.S. 106, Batavia, IL 60510}

\maketitle

\begin{abstract}
Perturbative coefficients for Wilson loops and the static-quark
self-energy are extracted from Monte Carlo simulations at weak
coupling. The lattice volumes and couplings are chosen to ensure
that the lattice momenta are all perturbative.
Twisted boundary conditions are used to eliminate the
effects of lattice zero modes and to suppress nonperturbative
finite-volume effects due to $Z(3)$ phases. Simulations of the
Wilson gluon action are done with both periodic and twisted
boundary conditions, and over a wide range of lattice volumes
(from $3^4$ to $16^4$) and couplings (from $\beta \approx
9$ to $\beta \approx 60$). A high precision comparison is made
between the simulation data and results from finite-volume
lattice perturbation theory. The Monte Carlo results are shown to
be in excellent agreement with perturbation theory through
second order. New results for third-order coefficients for a number
of Wilson loops and the static-quark self-energy are reported.
\end{abstract}

\section{Introduction}
\label{sec:Introduction}

Simulations using highly-improved lattice actions have become
commonplace in recent years. Effective use of these requires
perturbative matching calculations for masses, coupling constants
and currents, among other quantities. Higher-order perturbative
calculations for these actions are laborious but they are essential
in order to obtain precision results for most observables.

An alternative to doing calculations in analytical
perturbation theory is to directly measure short-distance quantities
in Monte Carlo simulations at weak coupling, as proposed in
Refs.\ \cite{DimmLat,DimmThesis}.
One exploits the fact that the lattice theory on a finite volume
enters a perturbative phase at weak coupling. In effect the couplings
and volumes in the simulations are chosen to ensure that the
lattice momenta are all perturbative (up to possible zero modes).
In this way one can in principle extract perturbative expansions for
many quantities, by fitting Monte Carlo data for appropriate
correlation functions to a power series in the coupling.

This approach has been shown to reproduce analytical results for the
first-order mass renormalization for Wilson fermions, and the first-order
additive self-energy for NRQCD fermions \cite{DimmLat,DimmThesis}.
In addition, preliminary estimates of some third-order Wilson loop
coefficients were made in Ref.\ \cite{DimmWilsonLoops}. An extension
of this technique to background field calculations was considered
in Ref.\ \cite{TrottierLepage}. Preliminary work on perturbative
simulations of quark actions has also been done \cite{Norm,Juge}.

The Monte Carlo method as implemented here requires as input from
conventional lattice perturbation theory only the expansion of the
average plaquette, and of the static potential (or some other quantity
that defines a physical coupling constant), to the desired order,
along with an estimate of the scale relevant to the quantity of
interest. These inputs are necessary because we use a renormalization
scheme defined through the perturbative expansion of the static-quark
potential \cite{LepMac,NRQCDalphas}. The renormalized coupling
$\alpha_P$ in this scheme can be extracted from measured values
of the average plaquette, given its perturbative expansion.

The method proceeds as follows. Simulations are done at a several
different values of $\beta$ ($\beta=2N/g^2$ for SU($N$) gauge theory).
At each $\beta$ we use the measured value of the plaquette to solve
for the value of the renormalized coupling $\alpha_P(q^*_{1,1})$,
at the scale $q^*_{1,1}$ that is
optimal for the plaquette \cite{LepMac}. We then run the couplings at
each $\beta$ to the scale $q^*$ appropriate to the quantity of
interest, whose expectation values are then fit to a truncated series
in $\alpha_P(q^*)$. The fit yields numerical values for the
perturbative coefficients.
To assess the effects of the truncation of the perturbation
series at finite order in $\alpha_P$, the fits are done including many
higher-order terms, beyond the order of interest, but where the fits
incorporate constraints on the coefficients \cite{PeterBayes},
which are required to lie in a range of values that is consistent
with a well-behaved perturbative expansion. One can also improve the
quality of the results by using lower-order coefficients from
conventional perturbation theory, if available, in order to further
constrain the fits to the Monte Carlo data, thereby obtaining more
accurate values for previously unknown higher-order terms.

In order to ensure that the lattice momenta are all sufficiently
perturbative, simulations are done on lattices for which
\begin{equation}
   {2 \pi \over La} \gg \Lambda_{\rm QCD} ,
\label{Lperturbative}
\end{equation}
where $\Lambda_{\rm QCD}$ is the QCD scale parameter, $a$ is the
lattice spacing, and $La$ is the physical length of the lattice.
In order to minimize perturbative finite-volume errors we must also
ensure that the spacing between lattice momenta is small compared
to the characteristic momentum scale $q^*$ associated with the
quantity of interest
\begin{equation}
   {2 \pi \over La} \ll q^* .
\label{Lfinitevolume}
\end{equation}

In practical simulations the lattices are such that
Eq.\ (\ref{Lperturbative}) is extremely well satisfied:
$a\Lambda_{\rm QCD}$ ranges from $10^{-3}$ to $10^{-29}$ in
the analyses in this paper. Moreover the quantities studied
here all have characteristic scales near the ultraviolet cutoff,
hence $a q^*$ is independent of $a$, and Eq.\ (\ref{Lfinitevolume})
can be easily satisfied. In theories with additional
scales, such as quark masses $m$, the characteristic scale $q^*$
is proportional to $f(ma)/a$, where $f$ is some dimensionless
function, hence $a q^*$ is independent of $a$ provided that
$m$ is adjusted to keep $ma$ fixed.
Perturbative finite-volume effects can also be analyzed in detail
by running simulations at several different volumes.

In simulations with periodic boundary conditions (PBC) however
there are lattice zero modes which will violate
Eq.\ (\ref{Lperturbative}). One must also ensure that nonperturbative
finite-volume effects arising from $Z(N)$ phases are sufficiently
suppressed. One way to achieve this, using a local updating algorithm,
is to work on lattices with sufficiently large volumes,
where the simulation is started with perturbative initial values
for the links (e.g., by setting all links to the identity). A more
powerful approach, analyzed in detail here, is to adopt boundary
conditions that eliminate zero modes and suppress transitions
between different phases.

It is also desirable to have a more general means for dealing
with potential infrared problems. The
effects of lattice zero modes on most observables is not known
and, while these could in some cases be accounted for by a
numerical fit to the data, it is preferable to eliminate these
states from the outset. More generally zero modes can significantly
alter the expected perturbative form of correlation functions.
For example, the presence of zero modes would pose a significant
problem for extracting quark masses, as one could not assume the
existence of an isolated pole in the perturbative quark propagator.
Twisted boundary conditions (TBC) \cite{tHooft,Luscher,Arroyo} can
be used to eliminate zero modes, and are easily incorporated
into simulations using existing code for a given action.

In this work we present a comprehensive study of this Monte Carlo
method for extracting perturbative quantities \cite{Lat99}.
We do simulations of SU(3) gauge theory using the Wilson gluon
action. The evolution of the $\alpha_P$ coupling is known for
this theory through three-loop order which, in principle, allows
one to use the method to determine perturbative expansions
through fourth order. Simulations are done for
both PBC and TBC. We show that using TBC is an effective
means of suppressing nonperturbative finite-volume effects due to
$Z(3)$ phases, as well as eliminating the effects of lattice
zero modes. We also make an extensive analysis of perturbative
finite-volume effects for TBC. We analyze simulation results for
a large set of Wilson loops and the static-quark self-energy, over
a wide range of couplings and lattice volumes. A high-precision
comparison is made between the simulation data and results from
finite-volume lattice perturbation theory. The Monte Carlo results
are shown to be in excellent agreement with perturbative
calculations through second order, which are available for
these observables for both periodic \cite{Heller,Martinelli}
and twisted \cite{OurTwistedPT} boundary conditions.
New results for third-order coefficients for fourteen different
Wilson loops and the static-quark self-energy are also reported.

Wilson loops provide a good quantity for a first test of the
Monte Carlo method, since small loops are relatively insensitive
to nonperturbative phases, and have small finite-volume errors.
The perturbative expansion of small Wilson loops is also relevant
to determinations of the strong coupling from lattice
simulations \cite{NRQCDalphas}.
The calculation of the static-quark self-energy is considerably
more involved, as it is very sensitive to nonperturbative phases and
has large perturbative finite-volume corrections. The static-quark
self-energy thus represents a good prototype for more realistic
calculations of other perturbative quantities, such as quark
masses. The self-energy is also useful for determinations of
the $b$-quark mass from lattice simulations \cite{E0mbSims}.

The rest of this paper is organized as follows.
In Sect.\ \ref{sec:WilsonLoops} we use PBC to study
Wilson loops at large $\beta$. Simulations are done on $16^4$
lattices at nine couplings. We evaluate Wilson loops of sizes
$R \times T$ with $R,T \leq 5$. The results are fit to a truncated
perturbation series, using the renormalized coupling $\alpha_P$
evaluated at a scale $q^*_{R,T}$ that is appropriate to each Wilson
loop \cite{LepMac}. We explicitly subtract the leading effects of
lattice zero modes from the Monte Carlo data using an existing
analytical calculation \cite{Coste}. A detailed comparison
is made with perturbation theory through second order,
and estimates are made of the third-order coefficients.

In Sect.\ \ref{sec:TwistedBC} we review how simulations are
done using twisted boundary conditions, and their use
in analytical perturbation theory. We demonstrate that
these boundary conditions can be used to virtually eliminate
nonperturbative effects due to $Z(3)$ phases even on lattices
with very small volumes.

The static-quark self-energy is analyzed in Sect.\ \ref{sec:Static}.
We extract the self-energy from the gauge-invariant Polyakov line,
which describes the propagation of a static quark across the entire
time-extent of the lattice. Other extractions of the self-energy
have relied upon large Wilson loops \cite{Martinelli,ParmaE0},
and the gauge-fixed quark propagator \cite{Norm} (the latter
methods could prove more efficient in Monte Carlo simulations,
as one can limit the propagation time on a lattice of a given
size). We make a perturbative analysis of the self-energy on
finite lattices, and show that one can sum leading logarithms
in the finite-volume correction. We present results of simulations
of the self-energy, with runs at nine volumes at each of nine
couplings. We compare the simulation results with perturbation
theory through second order, over the whole range of lattice
sizes studied here, and make an estimate of the third-order
self-energy, including an extrapolation to the infinite-volume
limit.

During the course of the calculations described in this paper,
we also investigated the question of how to design the most efficient
calculations through intelligent parameter choices, by using the
techniques of constrained curve fitting \cite{PeterBayes}.
In calculations like those in this paper, for example,
one can choose the couplings $\beta$ in order to minimize
the simulation cost required to achieve a given precision
in the perturbative coefficients. Although the Monte Carlo
calculations described in this paper were designed before
these optimization techniques were worked out, we include a
description of them here, in an Appendix, for use in future
calculations.

Some conclusions and prospects for future work are
briefly discussed in Sect.\ \ref{sec:Conclusions}.

\section{Wilson loops}
\label{sec:WilsonLoops}
In this section we analyze simulations of Wilson loops at large
$\beta$. The Wilson gauge-field action ${\cal S}_{\rm Wil}$ for
SU(3) color is used, where
\begin{equation}
   {\cal S}_{\rm Wil}[U] = \beta \sum_{x,\mu,\nu}
   \left(1 - \case13 \mbox{ReTr} \, U_{\mu\nu}(x) \right) ,
\label{WilsonAction}
\end{equation}
and $U_{\mu\nu}$ is the plaquette. Simulations were done on
$16^4$ lattices at nine couplings. Details of the simulation
parameters are given in Table~\ref{tab:PeriodicSimulations}.
Periodic boundary conditions were used here, in order to make a
direct comparison with the first- and second-order perturbative
coefficients calculated on finite lattices by Heller and
Karsch \cite{Heller}. We will also verify that the simulations
reproduce the effects of lattice zero modes to leading order.

\begin{table}[tb]
\begin{center}
\caption{Simulation parameters for Wilson loop measurements.
These simulations were all done on $16^4$ lattices with periodic
boundary conditions. The lattice coupling $\beta$ for each simulation
was determined from the bare coupling $\alpha_{\rm lat}$
by $\beta = 6 / (4\pi\alpha_{\rm lat})$.
The measured values of the average plaquette are shown, along with
the renormalized couplings $\alpha_P(3.40/a)$ and scale masses
$\Lambda_P$ extracted from Eqs.\ (\ref{lnW11}) and (\ref{alphaq}).
Ten configurations were skipped between measurements in all cases,
and the observables were computed by binning the measurements in
bin sizes of 50, which resulted in negligible autocorrelations at
all couplings.}
\label{tab:PeriodicSimulations}
\begin{tabular}{cccccc}
    $\alpha_{\rm lat}$
  & $\beta$
  & Measurements
  & $\langle \frac13 \mbox{Re-Tr} U_\Box \rangle$
  & $\alpha_P(3.40/a)$
  & $a\Lambda_P$ \\
\tableline
0.010  & 47.746  & 2320  & 0.957542(1)  \ & 0.01049  & $5.47 \times 10^{-23}$  \\
0.015  & 31.831  & 2459  & 0.935857(2)  \ & 0.01614  & $8.63 \times 10^{-15}$  \\
0.020  & 23.873  & 2112  & 0.913829(3)  \ & 0.02209  & $1.05 \times 10^{-10}$  \\
0.025  & 19.099  & 1558  & 0.891441(3)  \ & 0.02839  & $2.94 \times 10^{-8}$   \\
0.030  & 15.915  &  860  & 0.868599(9)  \ & 0.03510  & $1.25 \times 10^{-6}$   \\
0.035  & 13.642  &  746  & 0.845305(8)  \ & 0.04225  & $1.81 \times 10^{-5}$   \\
0.040  & 11.937  &  748  & 0.821472(9)  \ & 0.04992  & $1.34 \times 10^{-4}$   \\
0.045  & 10.610  &  500  & 0.797038(11)   & 0.05819  & $6.40 \times 10^{-4}$   \\
0.050  & 9.459   &  500  & 0.771872(11)   & 0.06719  & $2.24 \times 10^{-3}$   \\
\end{tabular}
\end{center}
\end{table}

We analyze the logarithm of the Wilson loop
\begin{equation}
   - {1\over 2 (R+T)} \ln W_{R,T}
   = \sum_n c_n \, \alpha_P^n (q^*_{R,T}) ,
\label{lnWRT}
\end{equation}
using a renormalized coupling $\alpha_P$ that is determined from
measured values of the plaquette, according
to \cite{LepMac,NRQCDalphas}
\begin{equation}
   -\ln W_{1,1} \equiv
   {4\pi \over 3} \alpha_P(3.40/a)
   \left[ 1 - 1.1909 \, \alpha_P \right] .
\label{lnW11}
\end{equation}
The coupling $\alpha_P$ is {\it defined\/} such that the logarithm of
the plaquette has no third- or higher-order terms in its perturbative
expansion \cite{NRQCDalphas}. Other quantities, of course, do have
higher-order terms when expressed as a series in $\alpha_P$.
One can express $\alpha_P(3.40/a)$ as a series in the bare lattice
coupling $\alpha_{\rm lat}$, using the third-order expansion of the
plaquette given in Ref.\ \cite{Alles}
\begin{equation}
   \alpha_P(3.40/a) = \alpha_{\rm lat}
   +  4.564 \, \alpha_{\rm lat}^2
   + 28.566 \, \alpha_{\rm lat}^3
   + O(\alpha_{\rm lat}^4) .
\label{aPtoa0}
\end{equation}
The large coefficients in this expansion are an artifact of
$\alpha_{\rm lat}$; using $\alpha_P$ eliminates large
renormalizations of the bare coupling. We also note that the
logarithm of the Wilson loop is better behaved perturbatively
than the Wilson loop itself, due to the exponentiation of the
perturbative perimeter law; this is also why we have divided by
the perimeter in defining the perturbative coefficients in
Eq.\ (\ref{lnWRT}).

The perturbation series for each Wilson loop is evaluated
using the renormalized coupling at a scale $q^*_{R,T}$ determined
according to the procedure of Ref.\ \cite{LepMac}; the scale
corresponds to the typical momentum carried by a gluon in the
leading-order diagram for a given quantity. The scales are given
in Table \ref{tab:qRT}. The couplings at these scales are evaluated
by first measuring the average plaquette in the simulation, and solving
Eq.\ (\ref{lnW11}) for $\alpha_P(3.40/a)$. We then evolve to the
scale appropriate to the quantity under consideration using the
universal second-order beta function, plus the third-order term
for $\alpha_P$,
with \cite{PDG}
\begin{eqnarray}
   \alpha_P(q) = & & {4\pi \over \beta_0 \ln(q^2/\Lambda_P^2)}
   \Biggl[ 1 - {\beta_1 \over \beta_0^2}
              {\ln\left[ \ln(q^2/\Lambda_P^2) \right]
               \over \ln(q^2/\Lambda_P^2)}
            + {\beta_1^2 \over \beta_0^4 \ln^2(q^2/\Lambda_P^2)}
\nonumber \\
   & & \times \left(
   \left( \ln \left[ \ln(q^2/\Lambda_P^2) \right] - {1\over2} \right)^2
   + {\beta_{2,P} \beta_0 \over \beta_1^2} - {5\over4} \right) \Biggr] .
\label{alphaq}
\end{eqnarray}
For our quenched simulations $\beta_0 = 11$, $\beta_1=102$, and
$\beta_{2,P} = \beta_{2,\overline{\rm MS}} + B_P$, where
$\beta_{2,\overline{\rm MS}} = 2857/2$ is the third beta function
coefficient in the $\overline{\rm MS}$ scheme, and where
$B_P=-147.57$ can be obtained from existing three-loop
calculations \cite{Alles,LuscheralphaMS}, as
described in Ref.\ \cite{NRQCDalphas}. The values of
$\alpha_P(3.40/a)$ and $\Lambda_P$ for our simulations are given
in Table~\ref{tab:PeriodicSimulations}.

\begin{table}[tb]
\begin{center}
\caption{Scale parameters for the couplings for various Wilson loops.}
\label{tab:qRT}
\begin{tabular}{cccc}
     Loop  &  $aq^*_{R,T}$  &  Loop  &  $aq^*_{R,T}$  \\
\tableline
  $1\times2$  & 3.07   & $2\times5$  & 2.46 \\
  $1\times3$  & 3.01   & $3\times3$  & 2.45  \\
  $1\times4$  & 2.96   & $3\times4$  & 2.38 \\
  $1\times5$  & 2.95   & $3\times5$  & 2.35  \\
  $2\times2$  & 2.65   & $4\times4$  & 2.30  \\
  $2\times3$  & 2.56   & $4\times5$  & 2.27  \\
  $2\times4$  & 2.49   & $5\times5$  & 2.23  \\
\end{tabular}
\end{center}
\end{table}

One can convert a perturbative expansion in $\alpha_{\rm lat}$
to one in $\alpha_P(q)$, at the scale appropriate to a
particular quantity, through third order, using
\begin{eqnarray}
   \alpha_{\rm lat} & = & \alpha_P(q)
   - \alpha_P^2(q) \left[
     {\beta_0 \over 4\pi} \ln\left({\pi \over a q} \right)^2
        + 4.702 \right]
   \nonumber \\
   & + & \alpha_P^3(q) \left\{
     \left[ {\beta_0 \over 4\pi} \ln\left({\pi \over a q} \right)^2
            + 4.702 \right]^2
     - {\beta_1 \over (4\pi)^2} \ln\left({\pi \over a q}\right)^2
     - 7.841 \right\}
   + O(\alpha_P^4) .
\label{alat3}
\end{eqnarray}
This connection can be obtained from the third-order expansion
of the plaquette in the bare coupling \cite{Alles},
which can be used to solve for $\alpha_{\rm lat}$ in terms
of $\alpha_P(3.40/a)$, given its definition in Eq.\ (\ref{lnW11});
one can then use a perturbative expansion of the evolution
equations \cite{Rodrigo} to eliminate $\alpha_P(3.40/a)$ in
favor of $\alpha_P(q)$. Equation (\ref{alat3}) extends the
second-order connection between $\alpha_{\rm lat}$ and
$\alpha_P(q)$ given in Ref.\ \cite{LepMac}.

The first- and second-order perturbative coefficients for the
Wilson loop were computed by Heller and Karsch \cite{Heller} for
an expansion in the bare lattice coupling. We convert this
expansion to a series in $\alpha_P(q^*_{R,T})$ using
Eq.\ (\ref{alat3}). These perturbative calculations were
done on finite lattices with periodic boundary conditions,
neglecting the contribution of lattice zero modes.
Our simulations of the Wilson
loops were also done with PBC but do contain the effects of zero
modes. In Sect.\ \ref{sec:TwistedBC} we will consider simulations
using twisted boundary conditions to eliminate these states. For
Wilson loops however we can make use of an analytical calculation
of the zero mode piece $c_1^{\rm zero}$ of the first-order
coefficient in Eq.\ (\ref{lnWRT}), due to
Coste {\it et al.} \cite{Coste}
\begin{equation}
   c_1^{\rm zero} = {4\pi(RT)^2 \over 9 (R+T) V} ,
\label{c1zero}
\end{equation}
where $V$ is the lattice volume. We will use this expression to
explicitly subtract the leading-order effects of the zero modes
from the Monte Carlo data. In the following we will show the
Monte Carlo data after first making this zero mode subtraction,
unless explicitly noted otherwise; hereafter we will use $c_n$ to
denote the (finite-volume) coefficients without zero mode
contributions.

We present Monte Carlo results for the $5\times5$ Wilson loop
in Fig.\ \ref{fig:WilsonK1}, in terms of the quantity
\begin{equation}
   \kappa_1^{\rm MC} \equiv {1 \over \alpha_P(q^*_{R,T})}
   \left[ - {1 \over 2(R+T)} \ln W_{R,T}^{\rm MC} \right] ,
\end{equation}
which should exhibit the limit $\kappa_1 \to c_1$ as $\alpha_P\to0$.
We extract estimates of the perturbative coefficients $c_n$
from the Monte Carlo data by fitting the results to the
series expansion Eq.\ (\ref{lnWRT}) where, to begin with, we treat
all coefficients $c_{n\ge1}$ as unknown. We will compare the
fit values for $c_1$ and $c_2$ with the results from analytical
perturbation theory, which provides a stringent test of the
Monte Carlo method.

An important aspect of the fitting procedure is how to reliably
account for the systematic error arising from the truncation of the
fit function at a finite order in $\alpha_P$. Including too few terms
in the expansion in $\alpha_P$ results in a poor fit to the data,
while including too many higher-order terms results in very poorly
constrained values for the lowest-order coefficients which should,
in fact, make the dominant contributions to the data. This situation
can be remedied by incorporating constraints on the coefficients,
which are required to lie in a range of values that is consistent
with our expectation that the perturbative expansion is well behaved.
We do this in practice by using conventional least-squares fitting
routines, where the $\chi^2$ is augmented according to:
\begin{equation}
   \chi^2(c_n) \rightarrow \chi^2_{\rm aug}(c_n) \equiv \chi^2(c_n)
   + \sum_n { (c_n - \bar c_n)^2 \over \bar \sigma_n^2} ,
\label{chisqaug}
\end{equation}
which tends to constrain the fit values for the $c_n$ to the
interval $\bar c_n \pm \bar \sigma_n$. This approach can be
motivated by Bayesian statistical analysis \cite{PeterBayes}.

If perturbation theory is reliable we expect the coefficients $c_n$ to
be of $O(1)$. We performed least-squares fits to Eq.\ (\ref{lnWRT}),
minimizing $\chi^2_{\rm aug}$ with $\bar c_n = 0$ and $\bar \sigma_n = 5$
for the first five orders in the expansion. The dashed line in
Fig.\ \ref{fig:WilsonK1} shows the results of the fit for the
$5\times5$ Wilson loop; note that the curvature in the Monte Carlo data
$\kappa_1^{\rm MC}$ shows the sensitivity of the simulations to the
third-order term in the perturbative series. The quality of the fits
is very good, with $Q$-values in excess of 50\%.

\begin{table}[t]
\caption{Monte Carlo results for the first three perturbative
coefficients for selected Wilson loops ($c_{1,2,3}^{\rm MC}$). The
results were obtained from a simultaneous fit to the coefficients,
as discussed in the text. The first- and second-order coefficients
from perturbation theory for the same size lattice are
also shown ($c_{1,2}^{\rm PT}$). The effects of zero modes are not
included in the perturbation theory values, and were removed at
leading order from the simulation data. Note that more accurate
results for $c_3^{\rm MC}$ for the full set of Wilson loops are
given in Table \ref{tab:Wilsonc3}, where the fits are done with
$c_1$ and $c_2$ constrained to their perturbative values.}
\label{tab:Wilsonc12}
\begin{tabular}{c|cc|cc|c}
     Loop &  $c_1^{\rm MC}$    &  $c_1^{\rm PT}$
          &  $c_2^{\rm MC}$    &  $c_2^{\rm PT}$
          &  $c_3^{\rm MC}$ \\
\hline
$1\times2$  & 1.2037(2) & 1.2039  & $-1.244(16)$ & $-1.260$ & 0.0(5) \\
$1\times3$  & 1.2587(2) & 1.2589  & $-1.185(19)$ & $-1.198$ & 0.4(5) \\
$2\times2$  & 1.4337(2) & 1.4338  & $-1.312(19)$ & $-1.323$ & 1.1(5) \\
$3\times3$  & 1.6089(3) & 1.6089  & $-1.218(24)$ & $-1.217$ & 2.5(6) \\
$4\times4$  & 1.7067(4) & 1.7067  & $-1.213(29)$ & $-1.210$ & 3.4(6) \\
$5\times5$  & 1.7693(6) & 1.7690  & $-1.201(40)$ & $-1.177$ & 4.3(7) \\
\end{tabular}
\end{table}

The measured values of $c_1$ and $c_2$ are in excellent agreement with
perturbation theory, as shown in Table \ref{tab:Wilsonc12}, with an
accuracy of a few parts in $10^4$ for the first-order coefficients
and a few parts in $10^2$ for the second-order coefficients.
The third-order coefficient can also be resolved, here with
almost no input from analytical perturbation theory.
The fit values are very stable to changes in the values of
$\bar c_n$ and $\bar \sigma_n$ used in Eq.\ (\ref{chisqaug}).

\begin{table}[t]
\caption{Monte Carlo results for $c_3$, where the first- and
second-order coefficients are constrained to their values from
perturbation theory.}
\label{tab:Wilsonc3}
\begin{tabular}{cccc}
     Loop  &  $c_3^{\rm MC}$
  &  Loop  &  $c_3^{\rm MC}$ \\
\tableline
  $1\times2$  &   0.43(9) \    &  $2\times5$  &   2.52(17) \\
  $1\times3$  &   0.66(11)      &  $3\times3$  &   2.53(15) \\
  $1\times4$  &   0.84(12)      &  $3\times4$  &   2.98(17) \\
  $1\times5$  &   0.94(14)      &  $3\times5$  &   3.26(19) \\
  $2\times2$  &   1.41(11)      &  $4\times4$  &   3.40(19) \\
  $2\times3$  &   1.91(13)      &  $4\times5$  &   3.71(21) \\
  $2\times4$  &   2.28(15)      &  $5\times5$  &   3.91(23) \\
\end{tabular}
\end{table}

Note that if the Monte Carlo data are fit with $c_1$ constrained
to its value from perturbation theory, then the errors on $c_2$ are
reduced by a factor of about three, with fit values in agreement
with perturbation theory within the reduced errors. Similarly,
we obtain more accurate results for $c_3$ by fitting
the Monte Carlo data with $c_1$ and $c_2$ constrained to
their perturbative values. We did fits to Eq.\ (\ref{lnWRT})
for the next three orders in the perturbative expansion, minimizing
$\chi^2_{\rm aug}$ using $\bar c_n = 0$ and $\bar \sigma_n = 5$
for $n=3,4,5$. The results for $c_3$ are given in
Table \ref{tab:Wilsonc3}, where the fit errors are seen to be
about 10\%.

It is also interesting to verify that the simulations reproduce the
leading effects of the zero modes. A convenient way to visualize
these effects is to plot the quantity
\begin{equation}
   \kappa_2^{\rm MC} \equiv {1 \over \alpha_P^2(q^*_{R,T})}
   \left[ - {1\over 2 (R+T)} \ln W_{R,T}^{\rm MC}
   - c_1 \alpha_P \right] ,
\label{kappa2}
\end{equation}
where the first-order coefficient is set to its perturbative value.
We plot $\kappa_2$ for the $5 \times 5$ loop in
Fig.\ \ref{fig:WilsonK2}, after subtracting the leading-order zero
mode term from the Monte Carlo data. We see that the data reproduce
the second-order coefficient from perturbation theory, with
$\kappa_2 \to c_2$ as $\alpha_P \to 0$. In Fig.\ \ref{fig:WilsonZMode}
we plot $\kappa_2$ but where the leading-order zero mode is {\it not\/}
subtracted from the data. We see evidence of singular behavior in
$\kappa_2$ at small coupling, indicating that the first-order term is
not completely removed from the Monte Carlo data when the zero mode
component is not treated. The dashed line in Fig.\ \ref{fig:WilsonZMode}
shows the results of a fit to Eq.\ (\ref{lnWRT}), taking account of
the leading zero mode contribution, where the term
$c_1^{\rm zero}/\alpha_P$ is included in the fit line. This shows
explicitly that the Monte Carlo data are sufficiently accurate to
reveal the small contribution from the zero modes, at sufficiently
small couplings.

We also present results for the residual
\begin{equation}
   \kappa_3^{\rm MC} \equiv {1 \over \alpha_P^3(q^*_{R,T})}
   \left[ - {1\over 2 (R+T)} \ln W_{R,T}^{\rm MC}
   - c_1 \alpha_P - c_2 \alpha_P^2 \right]
\label{kappa3}
\end{equation}
for the $5\times5$ loop in Fig.\ \ref{fig:WilsonK3}, which is
convenient for visualizing the sensitivity of the Monte Carlo
data to the third- and fourth-order terms.
The statistical errors in the Monte Carlo data are too large to
resolve $c_4$, although the best fits suggest that $c_4$
is of the same order as $c_3$ for all the Wilson loops analyzed here.

A potential complication in our analysis of $c_3$ is that we have only
corrected the Wilson loop data for the effects of zero modes to first
order. However we expect that the leading contribution from zero modes
that remains is of $O(\alpha_P^2 (RT)^2 / V)$ which, given the lattice
volume and the range of couplings analyzed here, should only be
comparable to terms of $O(\alpha_P^4$). In fact there is no visible
effect of zero modes beyond first order, within statistical errors;
this would should up as singular behavior in $\kappa_3$ at small
$\alpha_P$ (compare Fig.\ \ref{fig:WilsonK3} for $\kappa_3$ with
the two plots of $\kappa_2$ in
Figs.\ \ref{fig:WilsonK2} and \ref{fig:WilsonZMode}). We also
note that while we have extracted values for $c_3$ on a finite
lattice, the volume is large enough that the results should give
a good approximation to the coefficients on an infinite lattice
(with the corrections expected to be of $O(1/V)$).

A determination of higher order terms in the
expansion of the $1\times 1$ and $2 \times 2$ Wilson loops
has also been made in Ref.\ \cite{ParmaWilson}, using
numerical simulations of the Langevin equations, where a
perturbative expansion in the bare lattice coupling is applied
to the evolution equations themselves. The results are presented in
Ref.\ \cite{ParmaWilson} as an expansion of the Wilson loops
in powers of the bare lattice coupling,
$W_{R,T} = 1 - \sum_n \widetilde c_n \alpha_{\rm lat}^n$.
We can convert our third-order result for the expansion of
$\ln(W_{R,T})$ in $\alpha_P(q^*_{R,T})$, to an expansion
of $W_{R,T}$ in $\alpha_{\rm lat}$, by using the inverse
of Eq.\ (\ref{alat3}). We find
$\widetilde c_3 = 0.3 \pm 0.9$ for the $2\times 2$ Wilson loop,
in agreement with the result $\widetilde c_3 = 0.0 \pm 0.9$
reported in Ref.\ \cite{ParmaWilson}.

We note parenthetically that the expansion of the Wilson loop itself
is very poorly convergent, and that the vanishingly small
value of $\widetilde c_3$ for the $2 \times 2$ loop is accidental.
One finds very large expansion coefficients for other Wilson loops.
These expansions are tamed by taking the logarithm, and expressing
the series in $\alpha_P(q^*)$. For example, our
results give $\widetilde c_3 / \widetilde c_1 = 76.1 \pm 0.1$ for
the $5 \times 5$ loop, with this large value arising almost entirely
from the exponentiation of the perturbative perimeter law, and
the renormalization of the bare coupling (compare with
$c_3 / c_1 = 2.2 \pm 0.1$ for the logarithm of the $5 \times 5$
loop). In Sect.\ \ref{sec:Static} we show that the third-order
expansion of the static-quark self-energy is also very reasonable
when expressed in terms of $\alpha_P(q^*)$, but is very poorly
convergent when expressed in terms of $\alpha_{\rm lat}$.

\section{Twisted boundary conditions}
\label{sec:TwistedBC}

\subsection{Formalism}
\label{sec:TwistedFormalism}

The analysis of the preceding section shows that simulations at
large $\beta$ can be used to make accurate determinations of
perturbative quantities at higher orders than have been achieved
using conventional perturbation theory. However, as discussed in
Sect.\ \ref{sec:Introduction}, simulations with periodic boundary
conditions (PBC) are subject to the effects of lattice zero modes,
and a more convenient method for dealing with potential infrared
problems in more general situations is required. Fortunately
lattice zero modes can be completely eliminated by using twisted
boundary conditions (TBC). We also find that TBC significantly
reduce nonperturbative finite-volume effects due to $Z(N)$ phases.

Twisted boundary conditions \cite{tHooft,Luscher,Arroyo} for the
link fields resemble a gauge transformation on fields which cross
through selected lattice boundaries
\begin{equation}
   U_\alpha(x + L \hat\nu) = \Omega_\nu U_\alpha(x) \Omega_\nu^\dagger ,
\label{Utwist}
\end{equation}
where we take the twist matrices $\Omega_\nu$ to be constant.
A twist must be applied to at least two boundaries $\mu$ and $\nu$,
as the transformation matrix $\Omega_\nu$ for a single boundary can
be completely eliminated by a field redefinition.
The requirement that $U_\alpha(x + L \hat\mu + L \hat\nu)$,
which can be connected to $U_\alpha(x)$ by crossing the two lattice
boundaries in different orders, be single-valued implies that
the twist matrices must satisfy the algebra
\begin{equation}
   \Omega_\mu \Omega_\nu = \eta \Omega_\nu \Omega_\mu,
   \quad \eta \in Z(N) ,
\label{Omega}
\end{equation}
where we consider the general SU($N$) gauge theory in most of this
section. A pair of twist matrices generates a multiplication table
that forms a discrete subgroup of SU($N$). In particular \cite{Luscher}
\begin{equation}
   \Omega_\nu^N = (-1)^{N-1} I,
\label{OmegaN}
\end{equation}
where $I$ is the unit matrix.

The Wilson gluon action for TBC is written in terms of link variables
in the usual way, Eq.\ (\ref{WilsonAction}). The matrices
$\Lambda(x)$ that generate gauge transformations
\begin{equation}
   U_\mu(x) \rightarrow \Lambda(x) U_\mu(x) \Lambda^\dagger(x+\hat\mu)
\end{equation}
are subject to the same TBC as the links (Eq.\ (\ref{Utwist})).
One consequence of this is that the Polyakov line $P_\mu$ in a
twisted direction $\mu$ must have an additional factor of the
corresponding twist matrix, if it is to be gauge-invariant:
\begin{equation}
   P_\mu \equiv \left\langle
   U_\mu(x) U_\mu(x+\hat\mu) \ldots
   U_\mu(x+L\hat\mu) \times \Omega_\mu \right\rangle.
\label{Poly}
\end{equation}
The only zero-action fields are pure-gauge configurations
$U_\mu(x) = \Lambda(x) \Lambda^\dagger(x+\hat\mu)$ \cite{Luscher},
including possible $Z(N)$ phases. The action also possesses a
discrete symmetry
\begin{equation}
   U_\alpha(x) \rightarrow \Omega_i U_\alpha \Omega_i^\dagger
\label{TBCsymmetry}
\end{equation}
where $\Omega_i = \Omega_\mu, \Omega_\nu, \Omega_\mu\Omega_\nu$, etc.
Note that Eq.\ (\ref{TBCsymmetry}) is not a gauge transformation,
because $\Lambda(x)=\Omega$ does does not satisfy TBC \cite{Luscher}.
This symmetry implies that the Polyakov lines in the twisted
directions have zero expectation value, even in the perturbative
phase of the theory.

We have done SU(3) simulations with twisted boundary conditions
across two ``spatial'' boundaries $x$ and $y$
\begin{equation}
   \Omega_x \Omega_y =  \eta \Omega_y \Omega_x \quad (\mbox{T}xy) ,
\end{equation}
where $\eta = e^{2\pi i/3}$, and across all three spatial boundaries
$x$, $y$ and $z$
\begin{eqnarray}
   \Omega_x \Omega_y & = & \eta \Omega_y \Omega_x,
   \nonumber \\
   \Omega_x \Omega_z & = & \eta \Omega_z \Omega_x \quad (\mbox{T}xyz) .
\end{eqnarray}
We refer to these two cases as T$xy$ and T$xyz$ boundary conditions,
respectively.

Explicit representations of the twist matrices $\Omega_\nu$
are not needed since, as shown in Ref.\ \cite{Luscher}, one can absorb
them by a field redefinition of the link variables, leaving only phase
factors $\eta$ and $\eta^*$ multiplying the plaquettes at the
corners of the twisted planes. We have done simulations with this
field redefinition, and also using an explicit representation of the
boundary conditions Eq.\ (\ref{Utwist}) using the matrices (for SU(3))
\begin{equation}
   \Omega_x = \left[ \begin{array}{ccc}
                         0 & 1 & 0 \\
                         0 & 0 & 1 \\
                         1 & 0 & 0 \\
                         \end{array} \right] , \quad
   \Omega_y = \left[ \begin{array}{ccc}
                         e^{-2\pi i/3} & 0  &   0   \\
                         0             & 1  &   0   \\
                         0             & 0  & e^{2\pi i/3} \\
                         \end{array} \right] ,
\end{equation}
and
\begin{equation}
   \Omega_z = \Omega_y \Omega_x^2
            = \left[ \begin{array}{ccc}
                         0  &  0             & e^{-2\pi i/3} \\
                         1  &  0             & 0 \\
                         0  & e^{2\pi i/3}   & 0\\
              \end{array} \right] .
\end{equation}

The boundary conditions Eq.\ (\ref{Utwist}) lead to an unusual
quantization of the lattice momentum modes, as well as removing
the zero modes. Making the usual substitution
\begin{equation}
   U_\mu(x) = e^{i g a A_\mu(x)}
\end{equation}
the boundary conditions take the form
$A_\mu(x+L\hat\nu) = \Omega_\nu A_\mu(x) \Omega_\nu^\dagger$.
A twisted plane wave basis is used to Fourier analyze the fields
\begin{equation}
   A_\mu(x) = {1 \over V N}
              \sum_k \chi_k
              \Gamma_k e^{i k \cdot (x + \frac12 \hat \mu)}
              \tilde A_\mu(k) ,
\label{Aplanewave}
\end{equation}
where, in order to obey the boundary conditions, the matrices
$\Gamma_k$ must satisfy the algebra
\begin{equation}
   \Omega_\nu \Gamma_k \Omega_\nu^\dagger = e^{i k_\nu L} \Gamma_k ,
\label{OmegaQuantize}
\end{equation}
and where $\chi_k$ enforces a constraint on the mode sum, to be
developed below (see Eqs.\ (\ref{chik}) and (\ref{nz})).

The quantization conditions follow by iterating
Eq.\ (\ref{OmegaQuantize}) $N$ times and using Eq.\ (\ref{OmegaN}).
One finds that momenta in twisted directions are quantized as if
the SU($N$) fields live on a lattice of length $L \times N$, rather
than the actual length $L$ (although some modes are excluded)
\begin{equation}
   k_\nu = \left\{ \begin{array}{cl}
   \displaystyle
   {2\pi \over L N} n_\nu , & \nu = \mbox{twisted direction}, \\
   \\
   \displaystyle
   {2\pi \over  L} n_\nu, & \nu = \mbox{periodic direction} .
   \end{array} \right.
\label{kQuantize}
\end{equation}
The extra momentum degrees-of-freedom come about because the
color structure of each mode is unique, up to a phase.
Substituting $\Gamma_k = \Omega_x^\alpha \Omega_y^\beta$
into Eq.\ (\ref{OmegaQuantize}) one finds, with a convenient
choice of phase \cite{Luscher},
\begin{equation}
   \Gamma_k = \Omega_x^{-n_y} \, \Omega_y^{n_x} \,
   \eta^{\case12 (n_x + n_y) (n_x + n_y - 1)} .
\end{equation}
These matrices are orthonormal under the trace
\begin{equation}
  {1 \over N} \mbox{Tr}\left( \Gamma_{k'}^\dagger \Gamma_k \right)
  = \left\{ \begin{array}{cl}
            1 & \ \mbox{if\ } n'_{x,y} = n_{x,y} \ \mbox{mod\ }(N) , \\
            0 & \ \mbox{otherwise} .
    \end{array} \right.
\end{equation}
Since the fields $A_\mu$ must be traceless one finds that a
set of modes, including the zero modes, are excluded
\begin{equation}
   \chi_k = \left\{
   \begin{array}{cl}
   0 & \ \mbox{if\ } n_x = n_y = 0 \ ({\rm mod\ } N) , \\
   1 & \ \mbox{otherwise} .
   \end{array}
   \right.
\label{chik}
\end{equation}
In the case of T$xyz$ boundary conditions, a further constraint
emerges from Eq.\ (\ref{OmegaQuantize})
\begin{equation}
   \chi_k = \left\{
   \begin{array}{cl}
   1 & \ \mbox{if\ } n_z = 2n_x + n_y \ ({\rm mod\ } N) , \\
   0 & \ \mbox{otherwise} .
   \end{array}
   \right.
   \quad (\mbox{T}xyz)
\label{nz}
\end{equation}
Hence there is a factor of $N^2-1$ more momentum modes with TBC,
each of which has a single color degree-of-freedom, which is
exactly the number of independent colors that one has
for each momentum mode with PBC.

At tree-level the twisted gluon propagator in momentum space has
the structure
\cite{Luscher}
\begin{equation}
\label{Apropagator}
   \langle \tilde A_\mu(k) \tilde A_\nu(k') \rangle_{g=0} =
   \case12 V N \chi_k \eta^{-\case12 (k',k)} \delta_{k,k'}
   D_{\mu\nu}(k) ,
\end{equation}
where
\begin{equation}
   (k',k) \equiv n_x' n_x + n_y' n_y + (n_x + n_y) (n_x' + n_y') .
\end{equation}
For the Wilson action in Lorentz gauges one has
\begin{equation}
   D_{\mu\nu}(k) = {1 \over \hat k^2} \left[
   \delta_{\mu\nu} -
   (1 - \alpha) {\hat k_\mu \hat k_\nu \over \hat k^2} \right] ,
\label{Dmunu}
\end{equation}
with $\hat k_\mu = 2 \sin(\case12 k_\mu)$ and
$\hat k^2 = \sum_\lambda \hat k_\lambda^2$.

\subsection{Suppression of $Z(N)$ phases}
\label{sec:TwistedTunneling}

The Polyakov line along an untwisted direction is an order
parameter for the $Z(N)$ degenerate vacua of the lattice theory,
which correspond to the invariance of the Wilson action
under the transformation
\begin{equation}
   U_\mu(x) \to \eta U_\mu(x),
   \quad \forall x \ni x \cdot \hat\mu = \mbox{constant} .
\end{equation}
The Polyakov line is also sensitive to the formation of domains
between different $Z(N)$ phases. These nonperturbative effects must
be suppressed if one is to use Monte Carlo simulations at large
$\beta$ to extract perturbative quantities. In particular we will
use simulation results for the Polyakov line itself to obtain
the perturbative self-energy of a static quark.

In order to suppress nonperturbative $Z(N)$ phases we start the
simulation with all links initialized to a ``cold start,'' $U_\mu = I$.
The probability of making a transition to an nontrivial $Z(N)$ phase
in a local updating algorithm can then be reduced by working at
sufficiently large lattice volumes. In fact our results for Wilson loops
on $16^4$ lattices with PBC, presented in Sect.\ \ref{sec:WilsonLoops},
are in excellent agreement with finite-volume perturbation theory.
However we find that nonperturbative $Z(N)$ phases are generated
frequently on small lattices when PBC are used, and this occurs
even at extremely large $\beta$. On the other hand, we find that using
TBC leads to a dramatic suppression of these effects compared to PBC,
on lattices of the same size.

We illustrate the effects of $Z(3)$ phases with simulation results for
$4^4$ lattices at $\beta=9$. We show run time histories and scatter
plots of the real and imaginary parts of the Polyakov line along an
untwisted direction (hereafter taken to be the ``temporal''
direction $t$), where
\begin{equation}
   P_t(L) \equiv {1 \over 3 L^3} \sum_{\vec x}
   \mbox{ReTr} \left\langle \prod_{x_t=1}^L U_t(x) \right\rangle,
\label{Polyakov4}
\end{equation}
for a lattice of volume $L^4$. Results for PBC are shown in
Fig.\ \ref{fig:TunnelPBC}, for T$xy$ boundary conditions in
Fig.\ \ref{fig:TunnelTxy}, and for T$xyz$ boundary conditions in
Fig.\ \ref{fig:TunnelTxyz}. We see that nonperturbative
$Z(3)$ phases and domains render simulations with PBC useless
for extracting perturbative quantities on small lattices.
We also see that twisted boundary conditions create a barrier
between $Z(3)$ phases, and that transitions between these phases
and are essentially eliminated with T$xyz$ boundary conditions (with
no tunneling events observed in millions of updates in the range of
$\beta$ values considered here). In Sect.\ \ref{sec:Static} we
show that the remaining finite-volume effects on lattices with
T$xyz$ boundary conditions are very well described by perturbation
theory for $\beta \gtrsim 9$, even on volumes as small as $3^4$.

\section{Static-quark self-energy}
\label{sec:Static}

\subsection{Perturbation theory}
\label{sec:StaticPT}

In this section we consider the perturbative expansion of the
self-energy $E_0$ of a static quark. We extract the self-energy
from the gauge-invariant Polyakov line $P_t$ along an untwisted
direction, which describes the propagation of a static quark across
the entire time-extent of the lattice. One could also obtain the
self-energy from large Wilson loops \cite{Martinelli,ParmaE0}, or
from the gauge-fixed static-quark propagator \cite{Norm}. This
study however represents a good prototype for calculations of other
more realistic perturbative quantities, such as quark masses.

We first define the self-energy $E_0(L)$ on a finite lattice
according to
\begin{equation}
  a E_0(L) \equiv -{1 \over L} \ln(P_t(L))
\label{E0L}
\end{equation}
where, for comparison with our simulation results in the
next subsection, we consider lattices with equal lengths $L$
along all sides. One then obtains the infinite-volume self-energy
$E_0$ by taking the limit,
\begin{equation}
   E_0 = E_0(L \to \infty) .
\label{E0limit}
\end{equation}

We analyze the tadpole-improved self-energy. This is obtained
by dividing the links in the Polyakov line by a mean field
$u_0$, $U_\mu(x) \to U_\mu(x) / u_0$. Hence the tadpole-improved
self-energy is related to the unimproved self-energy by the
addition of $\ln(u_0)$. We use the average
plaquette to estimate the mean-field:
\begin{equation}
   u_0 = \langle U_\Box \rangle^{1/4}.
\label{u0}
\end{equation}

The expansion of the self-energy to second order was computed in
perturbation theory according to Eqs.\ (\ref{E0L}) and (\ref{E0limit})
by Heller and Karsch, for an expansion in the bare coupling,
with the result \cite{Heller,Martinelli}
\begin{equation}
   a E_0^{\rm unimp} =  2.1173 \, \alpha_{\rm lat}
       + 11.152  \, \alpha_{\rm lat}^2
       + O(\alpha_{\rm lat}^3)
\label{E0unimp}
\end{equation}
for the unimproved self-energy. Hereafter we consider only the
tadpole-improved self-energy, which we denote by $E_0$. We
convert Eq.\ (\ref{E0unimp}) to an expansion in the renormalized
coupling at the appropriate scale using Eq.\ (\ref{alat3}):
\begin{equation}
   a E_0 = 1.0701 \, \alpha_P(q^*_{E_0})
       + 0.117  \, \alpha_P^2
       + O(\alpha_P^3) ,
       \quad q^*_{E_0} = 0.84/a ,
\label{E0renPT}
\end{equation}
where Eq.\ (\ref{lnW11}), with couplings evolved
to $q^*_{E_0}$, provides the tadpole subtraction.

In the next subsection we will compare Monte Carlo data for the
self-energy with results from analytical perturbation theory on finite
lattices. In order to extract the infinite-volume self-energy $E_0$
from the Monte Carlo simulations we must also make an extrapolation
of measurements of $E_0(L)$ done on finite volumes. We can gain some
insight into the nature of the perturbative finite-volume corrections
from some analytical considerations.

We define perturbative coefficients $c_n(L)$ on a finite lattice
according to
\begin{equation}
   a E_0(L) = \sum_n c_n(L) \alpha_P^n(q^*_{E_0}) .
\label{E0Lseries}
\end{equation}
For TBC the first-order term is given by
\begin{equation}
  c_1(L) = {\pi \over N L^3}
  \sum_{\vec k} \chi_k D_{44}(k_4=0, \vec k) - {\pi \over 3}
  \quad [\mbox{TBC}] ,
\label{c1LTBC}
\end{equation}
to be compared with a calculation for PBC, ignoring the contribution
from zero modes
\begin{equation}
  c_1(L) = {\pi (N^2 - 1) \over N L^3}
  \sum_{\vec k \neq 0} D_{44}(k_4=0, \vec k) - {\pi \over 3}
  \quad [\mbox{PBC}] .
\label{c1LPBC}
\end{equation}
The constant $\pi/3$ that is subtracted from the momentum sums in the
above expressions is the value of the one-loop tadpole-improvement
counterterm, neglecting its very weak dependence on the lattice volume.
We remind the reader that the mode sums in Eqs.\ (\ref{c1LTBC})
and (\ref{c1LPBC}) are different, due to the different quantization
of the momentum components along the twisted and periodic
directions. The two sets of boundary conditions yield identical
results in the infinite-volume limit, where the color factor
$N^2 - 1$ emerges in the case of TBC because $\chi_k$ averages
to $N^2 - 1$ over infinitesimal momentum intervals.

Results for $c_1(L)$ for the three boundary conditions are
presented in Fig.\ \ref{fig:c1LPT}, which shows that finite-volume
effects are reduced with TBC, as suggested by Eq.\ (\ref{kQuantize}).
As is evident from the plots, finite-volume corrections are very
well parameterized by a simple linear form in $1/L$,
\begin{equation}
   c_1(L) = c_1 - X_1 {1 \over L} + O\left( {1\over L^2} \right) ,
\label{c1Lfitform}
\end{equation}
where $c_1 \equiv c_1(L=\infty)$. One can evaluate $X_1$ numerically
from Eqs.\ (\ref{c1LTBC}) and (\ref{c1LPBC}) with the results
$X_1 \approx 1.891$ (PBC), $0.254$ (T$xy$), and $0.771$ (T$xyz$).

We expect the finite-volume correction to run with a coupling
$\alpha_P(q^*_L)$, evaluated at an infrared scale $q^*_L$ that
is set by the box size:
\begin{equation}
   a E_0(L) = a E_0 - X_1 {\alpha_P(q^*_L) \over L}
   + O\left( {\alpha_P^2 \over L} , {\alpha_P \over L^2} \right) ,
   \quad q^*_L \propto {1 \over L} .
\label{VintCoulomb}
\end{equation}
A physical interpretation of this functional form for $c_1(L)$ is
that the static quark experiences a perturbative Coulomb interaction
with its images in the walls of the lattice.
The different values of the coefficient $X_1$ for different
boundary conditions also has a natural interpretation in this picture:
TBC reduce finite-volume effects by effectively putting the image
charges further away from the source charge.

Having established Eq.\ (\ref{VintCoulomb}), one can deduce
logarithms in $L$ in the self-energy at higher orders. For
example, at second order one has
\begin{equation}
   c_2(L) = c_2
   - {1 \over L} \left( X_2 + Y_2 \ln(L^2) \right)
   + O\left( {\ln(L^2) \over L^2} \right) ,
\label{c2Lfit}
\end{equation}
where $c_2 \equiv c_2(L=\infty)$, and
\begin{equation}
   Y_2 = X_1 {\beta_0 \over 4\pi} .
\label{Y2X1}
\end{equation}
This follows from an expansion of the running coupling in
Eq.\ (\ref{VintCoulomb}), to second order in the coupling at a
reference scale, such as $\alpha_P(q^*_{E_0})$.
One can explicitly isolate the logarithm in the
second order coefficient using existing perturbative calculations,
which were done long ago by Heller and Karsch in the case of
PBC \cite{Heller}, and which have also recently been done
in Ref.\ \cite{OurTwistedPT} for TBC.
Results of the perturbative calculations for the three boundary
conditions are plotted in Fig.\ \ref{fig:c2LPT} over a range of
lattice sizes. The dashed lines in Fig.\ \ref{fig:c2LPT} show
fits to Eq.\ (\ref{c2Lfit}), where $c_2$ is constrained to the
correct value; the fits are in excellent agreement
with Eq.\ (\ref{Y2X1}). Note that the curvature in the results
for $c_2(L)$ reveals the presence of the logarithm, particulary
in the case of PBC and T$xyz$ boundary conditions, where the
logarithm makes a significant contribution at the lattice sizes
shown in Fig.\ \ref{fig:c2LPT}.

In the next subsection we will use Eq.\ (\ref{VintCoulomb}) to deduce
the form of the logarithms in $L$ in the third order self-energy,
which will help to constrain the infinite-volume extrapolation of
the Monte Carlo data. We note that one should similarly be able to
determine the leading logarithms in the finite-volume corrections to
other quantities, which should likewise prove useful in
Monte Carlo determinations of their perturbative expansions.

\subsection{Self-energy from Monte Carlo simulations}
\label{sec:StaticMC}

We measured the static-quark self-energy in simulations done with
T$xyz$ twisted boundary conditions at nine couplings. The simulation
parameters are given in Table \ref{tab:TwistedSimulations}.
Simulations were run on nine volumes $L^4$, $L=[3,11]$ inclusive,
at each of the nine couplings, for a total of 81 lattices.
The number of measurements made on each volume, at all couplings
except $\beta=60$, were as follows: $2000$ measurements for the
lattices with $L=[3,6]$ inclusive, $1500$ measurements for $L=7$,
$1200$ for $L=8$, $800$ for $L=9$, $600$ for $L=10$, and
$400$ for $L=11$ (ten times as many measurements were made on
each volume at $\beta=60$). One hundred configurations were skipped
between measurements at all couplings, except at $\beta=60$, where
ten configurations were skipped between measurements. The
observables were computed by binning the measurements in bin sizes
of one hundred, which resulted in negligible autocorrelations at
all couplings. The static energy and its error were computed from
the binned ensembles using a standard jackknife analysis.

\begin{table}[tb]
\begin{center}
\caption{Simulation parameters for static self-energy measurements
with T$xyz$ twisted boundary conditions. At each $\beta$ simulations
were run on nine volumes, $L=[3,11]$ inclusive. The measured average
plaquette on the lattices with $L=10$ are given, along with the scale
mass $\Lambda_P$ computed from Eqs.\ (\ref{lnW11}) and (\ref{alphaq}).
The bare coupling is shown along the renormalized coupling
$\alpha_P(q^*_{E_0})$ evaluated at the scale appropriate to the
self-energy with tadpole renormalization.}
\label{tab:TwistedSimulations}
\begin{tabular}{ccccc}
    $\beta$
  & $\langle \frac13 \mbox{Re-Tr} U_\Box \rangle$
  & $a\Lambda_P$
  & $\alpha_{\rm lat}$
  & $\alpha_P(0.84/a)$ \\
\tableline
60.0 & 0.966311(1)\ \  & $2.57 \times 10^{-29}$  & 0.008  & 0.00843 \\
23.8 & 0.913831(4)\ \  & $1.05 \times 10^{-10}$  & 0.020  & 0.02338 \\
19.0 & 0.891415(5)\ \  & $2.95 \times 10^{-8}$   & 0.025  & 0.03058 \\
16.0 & 0.869332(6)\ \  & $1.13 \times 10^{-6}$   & 0.030  & 0.03824 \\
13.6 & 0.845310(7)\ \  & $1.81 \times 10^{-5}$   & 0.035  & 0.04731 \\
12.0 & 0.822493(8)\ \  & $1.25 \times 10^{-4}$   & 0.040  & 0.05676 \\
10.6 & 0.797025(10)    & $6.41 \times 10^{-4}$   & 0.045  & 0.06844 \\
 9.5 & 0.771866(11)    & $2.24 \times 10^{-3}$   & 0.050  & 0.08138 \\
 9.0 & 0.756142(13)    & $4.29 \times 10^{-3}$   & 0.053  & 0.09032 \\
\end{tabular}
\end{center}
\end{table}

To demonstrate the reliability of the Monte Carlo method, we first use
the simulation results to estimate the first- and second-order
perturbative coefficients. In Fig.\ \ref{fig:StaticK1L} we
plot the quantity
\begin{equation}
   \kappa_1^{\rm MC}(L) = a E_0^{\rm MC}(L) \, / \, \alpha_P(q^*_{E_0})
\label{StaticK1}
\end{equation}
versus $\alpha_P(q^*_{E_0})$, for all values of $L$. The dashed lines
show the results of least-squares fits to Eq.\ (\ref{E0Lseries}),
minimizing $\chi^2_{\rm aug}$ (Eq.\ (\ref{chisqaug})) using
$\bar c_n = 0$ and $\bar \sigma_n = 5$ for the first five
orders in the expansion. The quality of the fits is very good
in most cases, with $Q$-values typically in excess of about 20\%,
although the lowest $Q$-value in the fits is 3\%.

We show the Monte Carlo results for $c_1(L)$ in Fig.\ \ref{fig:c1LMC},
where they are compared with finite-volume perturbation theory for
T$xyz$ boundary conditions, Eq.\ (\ref{c1LTBC}). The data agree
with perturbation theory within errors of only a few parts in $10^3$.

Monte Carlo results for $c_2(L)$ are shown in Fig.\ \ref{fig:c2LMC}.
In this case the fits to Eq.\ (\ref{E0Lseries}) were done with
$c_1(L)$ constrained to its perturbative value. We see that the
Monte Carlo simulations also reproduce the results of second order
T$xyz$ perturbation theory \cite{OurTwistedPT} over the full
range of lattice sizes, within errors that are as small
as a few parts in $10^2$ at several volumes.

The third-order term in the self-energy is not known from conventional
perturbation theory. We use our simulation results to estimate
$c_3(L)$, by redoing fits to Eq.\ (\ref{E0Lseries}) with
both $c_1(L)$ and $c_2(L)$ constrained to their perturbative values.
In order to determine the value of $c_3$ at infinite volume, we must
also account for the systematic error due to the extrapolation from
finite lattices. The leading finite-volume corrections come from
logarithms in $L$, which can be determined using renormalization-group
arguments (cf.\ Eqs.\ (\ref{VintCoulomb})--(\ref{Y2X1})).
Making use of these constraints considerably improves the
accuracy of the extrapolation to infinite volume.

We show the Monte Carlo data for the third-order coefficient
as a function of lattice size in Fig.\ \ref{fig:c3LMC}, after
subtracting the logarithms at $O(1/L)$; the data are
presented in terms of the residual
\begin{equation}
   \delta c_3(L) \equiv c_3(L)
 + X_1 {\beta_0^2 \over (4\pi)^2} \,
   {1 \over L} \ln^2\left({L^2 \over L_0^2}\right)
 + X_1 {\beta_1 \over (4\pi)^2} \,
   {1 \over L} \ln\left({L^2 \over L_0^2}\right) .
\label{deltac3}
\end{equation}
The scale length $L_0$ in the leading logarithm is determined from
second-order perturbation theory where, according to Eq.\ ({\ref{c2Lfit}),
$L_0 = \exp(-X_2/2Y_2)$. We evaluate the scale length from
a fit to the T$xyz$ perturbation theory results illustrated in
Fig.\ \ref{fig:c2LPT}, which gives $L_0 \approx 0.45$ for the
expansion in $\alpha_P(q^*_{E_0})$.

The Monte Carlo data are consistent with the logarithms
in Eq.\ (\ref{deltac3}), which are found to dominate
the extrapolation to the infinite-volume limit, even from
these relatively small lattices.
To extract the infinite-volume coefficient $c_3$ we fit the
remaining finite-volume corrections to the form
\begin{equation}
   \delta c_3(L) = c_3 + p_{1,0} {1 \over L}
          + \sum_{m\geq2}{1 \over L^m} \sum_{n=0}^2
            p_{m,n} \ln^n\left({L^2 \over L_0^2}\right) .
\label{c3Lfit}
\end{equation}
Figure \ref{fig:c3LMC} shows the results of a fit to
Eq.\ (\ref{c3Lfit}) where $\chi^2_{\rm aug}$ is minimized using
$\bar c_3 = \bar p_{m,n} = 0$,
and $\bar \sigma_{c_3} = \bar \sigma_{p_{m,n}} = 4$, for
$m=2,3$ and $n=0,1,2$ (and for $p_{1,0}$). The fit yields
\begin{equation}
   c_3^{\rm MC} = 3.56 \pm 0.50 \quad \mbox{(infinite-volume limit)} .
\label{c3MC}
\end{equation}
Changing the order of the expansion in $1/L$ in
Eq.\ (\ref{c3Lfit}) makes litle change in the fit value for $c_3$.
We conclude from these results that renormalized perturbation theory
for the tadpole-improved self-energy is well behaved through
third order, with the data in Fig.\ \ref{fig:c3LMC} clearly
demonstrating that $c_3$ is of $O(1)$.

An estimate of the third-order term in the expansion of the unimproved
self-energy, in the bare lattice coupling (Eq.\ (\ref{E0unimp})),
has recently been reported using numerical simulations of the
Langevin equations \cite{ParmaE0}. We can convert our result for
$c_3$ from an expansion in $\alpha_P(0.84/a)$ to one in
$\alpha_{\rm lat}$, using the inverse of Eq.\ (\ref{alat3}). We find
$c_{3,{\rm lat}}^{\rm MC} = 86.6 \pm 0.5$
(without tadpole improvement), in agreement with the value
$c_{3,{\rm lat}} = 86.2 \pm 0.6$ reported in Ref.\ \cite{ParmaE0}.
The bare coupling is clearly a very poor expansion parameter, with
96\% of $c_{3,{\rm lat}}$ being absorbed by renormalization when
a physical coupling is used.

We note that the simulations in Ref.\ \cite{ParmaE0} were done on much
larger lattices than were used here. We were able to extract $c_3$
from smaller lattices because the leading finite-volume corrections
were identified using renormalization-group methods.
This determination of the third-order self-energy involved a
modest computational effort. The entire set of simulations in the
present analysis required only the equivalent of about 150 days of
running on a single 1~GHz processor.

\section{Summary and Outlook}
\label{sec:Conclusions}
The results presented here demonstrate that higher-order perturbative
expansions are accessible in Monte Carlo simulations at large
$\beta$. An extensive theoretical analysis was presented together
with the results of numerical simulations of a large set of Wilson
loops and the static-quark self-energy. Twisted boundary conditions
were investigated as a means of eliminating zero modes and
suppressing nonperturbative finite-volume artifacts, and
an extensive analysis of perturbative finite-volume corrections
was made. Wilson loops provided a good quantity for a first test
of the Monte Carlo method, since small loops
are relatively insensitive to finite-volume effects.
The calculation of the static-quark self-energy was considerably
more involved, as it is very sensitive to nonperturbative phases and
has large perturbative finite-volume corrections. The static-quark
self-energy thus represents a good prototype for calculations of
other more realistic perturbative quantities, such as quark masses.

The simulation results were shown to reproduce perturbation theory on
finite lattices through second order to high precision, over a wide
range of lattice sizes and couplings. Monte Carlo results for
the fourteen smallest Wilson loops were found to agree
with perturbation theory within the errors, with an accuracy
of a few parts in $10^4$ for the first-order coefficients, and a
few parts in $10^2$ for the second-order terms. The Monte Carlo
results for the static-quark self-energy were found to agree with
finite-volume perturbation theory over the full range of lattice
sizes analyzed here, with an accuracy of a few parts in $10^3$ at
first order, and a few parts in $10^2$ at second order. This
precision was achieved with relatively little computational
effort. New estimates of third-order terms for the Wilson loops
and the static-quark self-energy were obtained to about 10\% accuracy.
Renormalization-group arguments were used to improve the quality
of the extrapolation of the self-energy to infinite volume.
The results demonstrate that renormalized perturbation theory
for Wilson loops and the self-energy is well behaved
through third order.

These methods can be directly applied to improved gluon actions,
and can be extended to quark actions. We have done some work on
large $\beta$ simulations for fermions in the nonrelativistic
formulation of QCD, extending the preliminary studies reported
in Refs.\ \cite{DimmLat,DimmThesis}. We find that simulations
of the additive energy and multiplicative mass renormalization
reproduce results of one-loop perturbation theory, and can
resolve the second-order terms in the expansion of these
quantities, over a wide range of bare quark masses \cite{Norm}.
Further work in this direction is in progress.

\acknowledgements

We are very grateful to Christine Davies for several crucial
discussions. We thank Urs Heller for generously providing us
with his programs for second-order coefficients. We also
thank Richard Woloshyn, Mark Alford, Ron Horgan, Massimo DiPierro
and Kent Hornbostel for helpful conversations. This work was
supported in part by the National Science Foundation, the
U.S. Department of Energy, and by the National Science and
Engineering Research Council of Canada.
HDT would also like to thank the United Kingdom Particle Physics
and Astronomy Research Council, and the physics department
of Cornell University, for support during part of this work.

\section*{Appendix: \\ Designing optimized Monte Carlo simulations}
In this Appendix we consider the question of how to design
the most efficient calculations through intelligent parameter
choices, by using the
techniques of constrained curve fitting \cite{PeterBayes}.
As discussed in the Introduction, one objective of this analysis,
in the context of short-distance Monte Carlo simulations like
those in this paper, is to choose the couplings $\alpha$ for
the simulations so as to minimize the cost required to achieve
a given precision in the perturbative coefficients.

As described in Ref.~\cite{PeterBayes} and in
Sect.~\ref{sec:WilsonLoops}, the effects of truncation errors in
fitting power series to Monte Carlo data may by estimated by
augmenting $\chi^2$ with a function that constrains parameters,
which are poorly determined statistically, to plausible values.
Consider for example the logarithm of a Wilson loop, denoted
by $W$, which has the expansion
\begin{equation}
   W(\alpha) =
   c_1 \alpha +c_2 \alpha^2 +c_3 \alpha^3 +\dots.
\label{Wseries}
\end{equation}
We will use $\alpha_i$ ($i=1,2,\ldots,n_\alpha$) to denote the
set of couplings at which the simulations are done. We may define
an augmented $\chi^2$ as in Eq.\ (\ref{chisqaug}),
\begin{equation}
   \chi^2(c_n) \rightarrow \chi^2_{\rm aug}(c_n)
   \equiv \chi^2(c_n) +
   \sum_n\frac{(c_n-\bar c_n)^2}{\bar\sigma_n^2},
\end{equation}
where the second term on the right tends to constrain poorly
determined parameters to the range $\bar c_n \pm \bar \sigma_n$,
based on our prior experience with the power series.  In the
examples in this Appendix, we will use $\bar c_n=0$ and
$\bar \sigma_n=1$.

A well designed calculation should minimize the errors in the final
results, for a given amount of computer time. The uncertainties in
the fit parameters $c_n$ are determined from the inverse of the
Hessian matrix, which we denote by $H_{mn}$, where
\begin{equation}
   H_{mn} \equiv \frac{1}{2}\frac{\partial^2\chi^2}
   {\partial c_m \partial c_n}.
\end{equation}
Then the uncertainty in $c_n$ is
\begin{equation}
   \delta c_n=H^{-1}_{nn}.
\end{equation}

In the case of Eq.\ (\ref{Wseries}), an explicit expression
for the Hessian matrix can be obtained
\begin{equation}
   H_{mn} = \sum_{i=1}^{n_\alpha}
   {\alpha_i^{m+n} \over \sigma^2_W(\alpha_i)}
   + {\delta_{mn} \over \bar \sigma^2_n} ,
\end{equation}
where $\sigma_W(\alpha)$ is the statistical error in $W(\alpha)$.
We note that a useful approximation for the statistical error
in the Wilson loop (or its logarithm) is
\begin{equation}
   \sigma_W(\alpha) \approx f \alpha,
\end{equation}
where $f\propto 1/\sqrt{\rm CPU\ time}$ and is independent of
$\alpha$.

The optimal selection of the values $\alpha_i$ at which the
simulations are to be done may be determined by numerically
minimizing the $\delta c_n$ with respect to the $\alpha_i$.
The optimal placement of $\alpha$'s for the first couple of
parameters may be guessed without doing much calculation.
For example, since relative statistical errors
are independent of $\alpha$, $c_1$ is obtained most accurately
by running at the smallest possible coupling $\alpha_1$
(avoiding round-off errors), thereby minimizing truncation errors
from higher-order terms in the series expansion. The smallest total
error in $c_2$ is then obtained by choosing a second coupling
$\alpha_2$ such that the statistical error in the slope of
$\widetilde W / \alpha$, which is $\sim f \alpha_2 / \alpha_2 =f$,
is equal to the truncation error, which is given by
$\bar\sigma_3 \alpha_2^3/\alpha_2=\bar\sigma_3\alpha_2^2$.
Explicit numerical minimization reproduces expectations for
these simple cases, but also produces optimized design parameters
for arbitrary numbers of points and allocations of CPU time
among them. The results of a three-point optimization
are illustrated in Fig.\ \ref{fig:Aalphas}. One sees that as
the CPU time is increased, smaller and smaller couplings
are obtained from the minimization calculations, though the
optimal $\alpha$'s do not fall quite as quickly as the fourth
root of CPU time (as would be expected in a two-point fit).

Perturbative series in lattice QCD are typically used in
nonperturbative Monte Carlo calculations with $\alpha$'s
in the range of about 0.15--0.30.
To illustrate the effects of optimization calculations on the
final result of a nonperturbative simulation, we consider the
application of the perturbative series Eq.\ (\ref{Wseries})
to a simulation with $\alpha=0.25$. The results are shown
in Fig.\ \ref{fig:Aerrors}. The errors coming from the lowest-order
coefficients are greatest when the short-distance
simulations are done with low statistics, but these errors
decrease most rapidly with CPU time; hence the order of the
term that contributes the largest error to the nonperturbative
quantity rises as a function of the CPU time of the
perturbative simulations.

Similar results, different in detail, are obtained
for other quantities. For example, for the third order
coefficient $\kappa_3$ for the $5\times 5$ Wilson loop,
shown in Fig.\ \ref{fig:WilsonK3}, the statistical errors
are $f/\alpha^2$ and the truncation error is $\sigma_c \alpha$.
The minimization formula gives the optimal placement for a
single-point simulation as
\begin{equation}
\alpha=\left(\frac{f}{\sigma_c}\right)^{1/3},
\end{equation}
as expected.

Constrained curve fitting formulas for parameter fitting
provide a concrete way of translating estimates of truncation
errors into designs of efficient computations. For one or two
parameters, they lead to exactly the same the same choices
of parameters that intuitive guesswork provides. However, they
also provide clear optimizations in the much more common
situations in which we have too many parameters to guess
about, or in which we are trying to design new runs to
improve the results from imperfectly designed initial runs.

\begin{figure}[htb]
\psfig{file=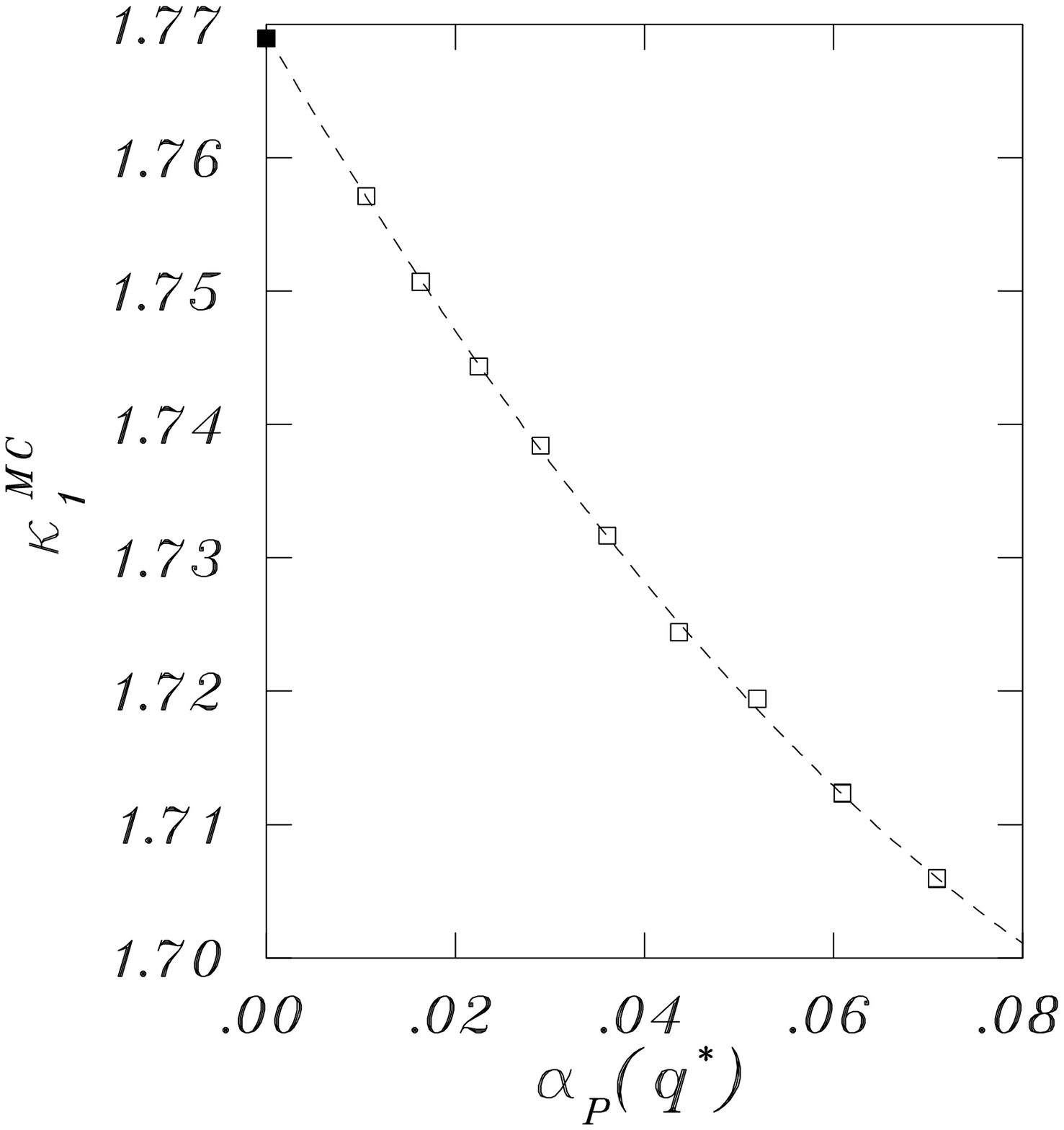,height=5.5in,width=5.in}
\vskip 0.25 in
\caption{Monte Carlo results for $\kappa_1^{\rm MC}$ for the
$5\times5$ Wilson loop, after the effects of zero modes are
removed at leading order from the simulation data using
Eq.\ (\ref{c1zero}). The statistical errors are smaller than
the plotting symbols. The filled square shows $c_1$ from
perturbation theory. The dashed line shows the results of a
fit to Eq.\ (\ref{lnWRT}).}
\label{fig:WilsonK1}
\end{figure}

\begin{figure}[htb]
\psfig{file=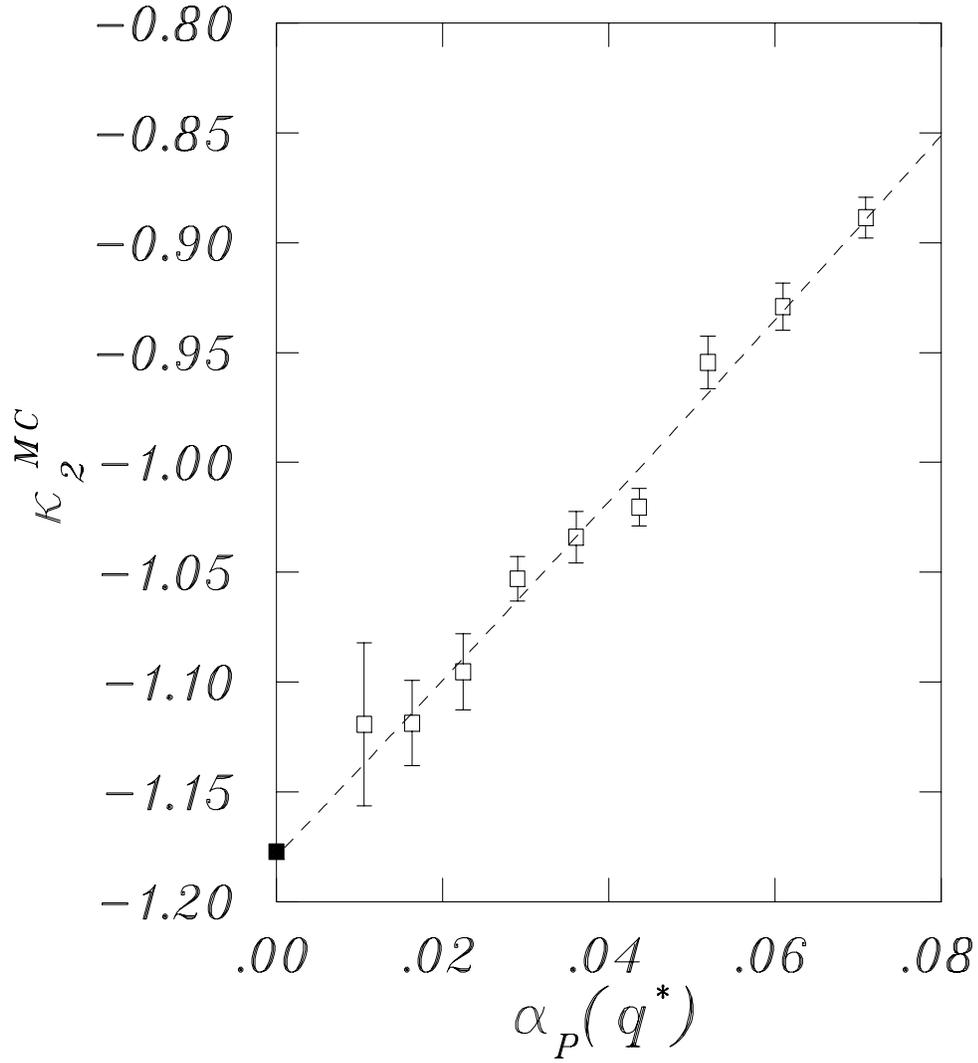,height=5.5in,width=5.in}
\vskip 0.25 in
\caption{Monte Carlo results for $\kappa_2^{\rm MC}$ for the
$5\times5$ Wilson loop, after the effects of zero modes are
removed at leading order from the simulation data. The filled
square shows $c_2$ from perturbation theory. The dashed line shows
the results of a fit to Eq.\ (\ref{lnWRT}), where $c_1$ is
constrained to its perturbative value.}
\label{fig:WilsonK2}
\end{figure}

\begin{figure}[htb]
\psfig{file=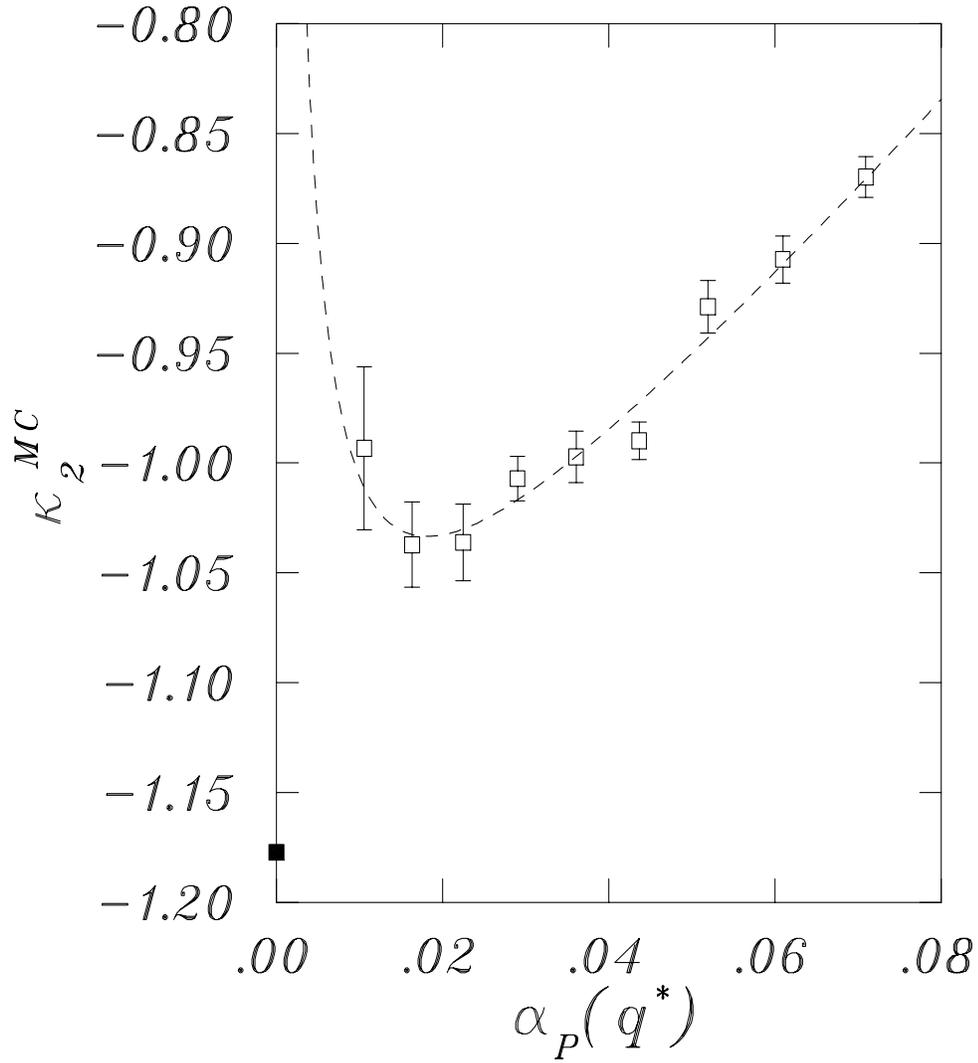,height=5.5in,width=5.in}
\vskip 0.25 in
\caption{Monte Carlo results for $\kappa_2$ for the $5\times5$
Wilson loop, when the effects of the first-order zero mode term
are {\it not} removed from the simulation data. The dashed line
shows the results of a fit to the data, described
in the text.}
\label{fig:WilsonZMode}
\end{figure}

\begin{figure}[htb]
\psfig{file=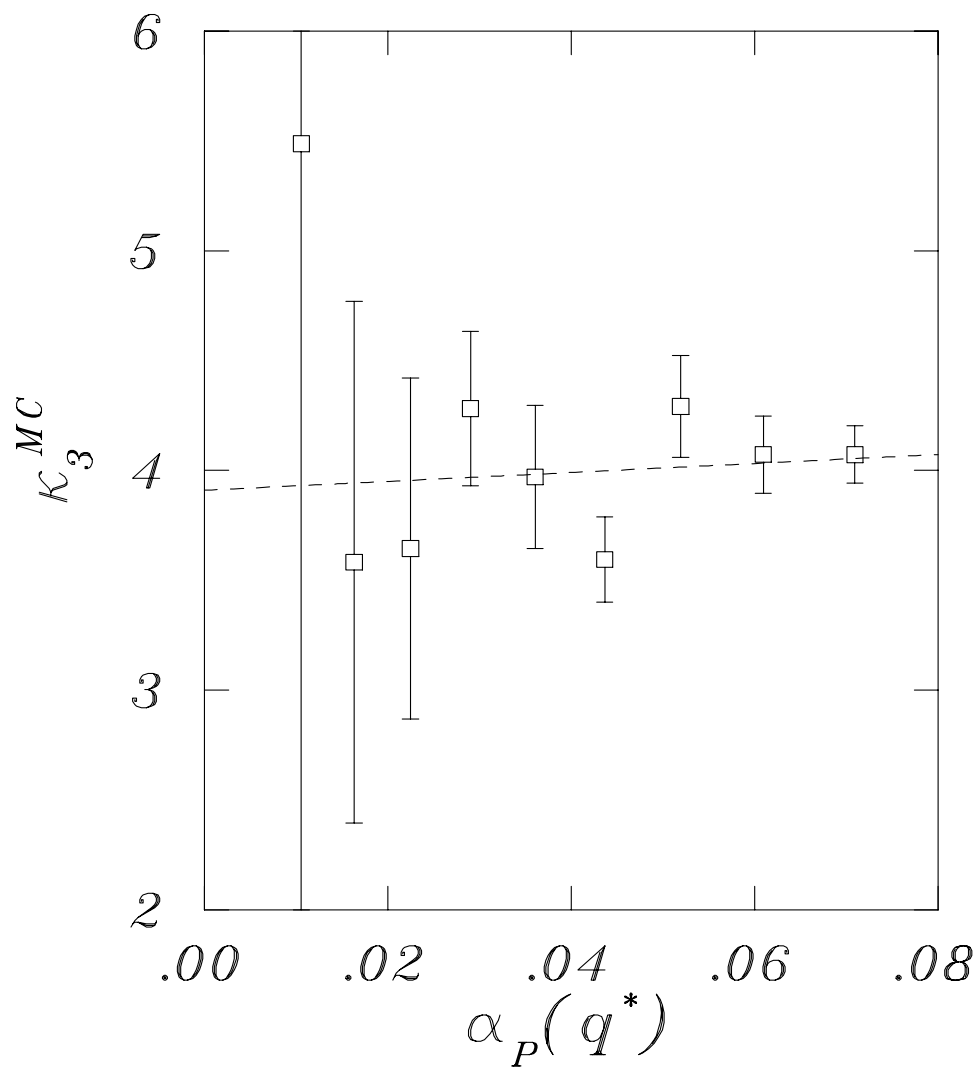,height=5.5in,width=5.in}
\vskip 0.25 in
\caption{Results for $\kappa_3$ for the $5\times5$ Wilson loop.
The dashed line shows the results of a fit to Eq.\ (\ref{lnWRT}),
where $c_1$ and $c_2$ are constrained to their perturbative values.}
\label{fig:WilsonK3}
\end{figure}

\begin{figure}[htb]
\psfig{file=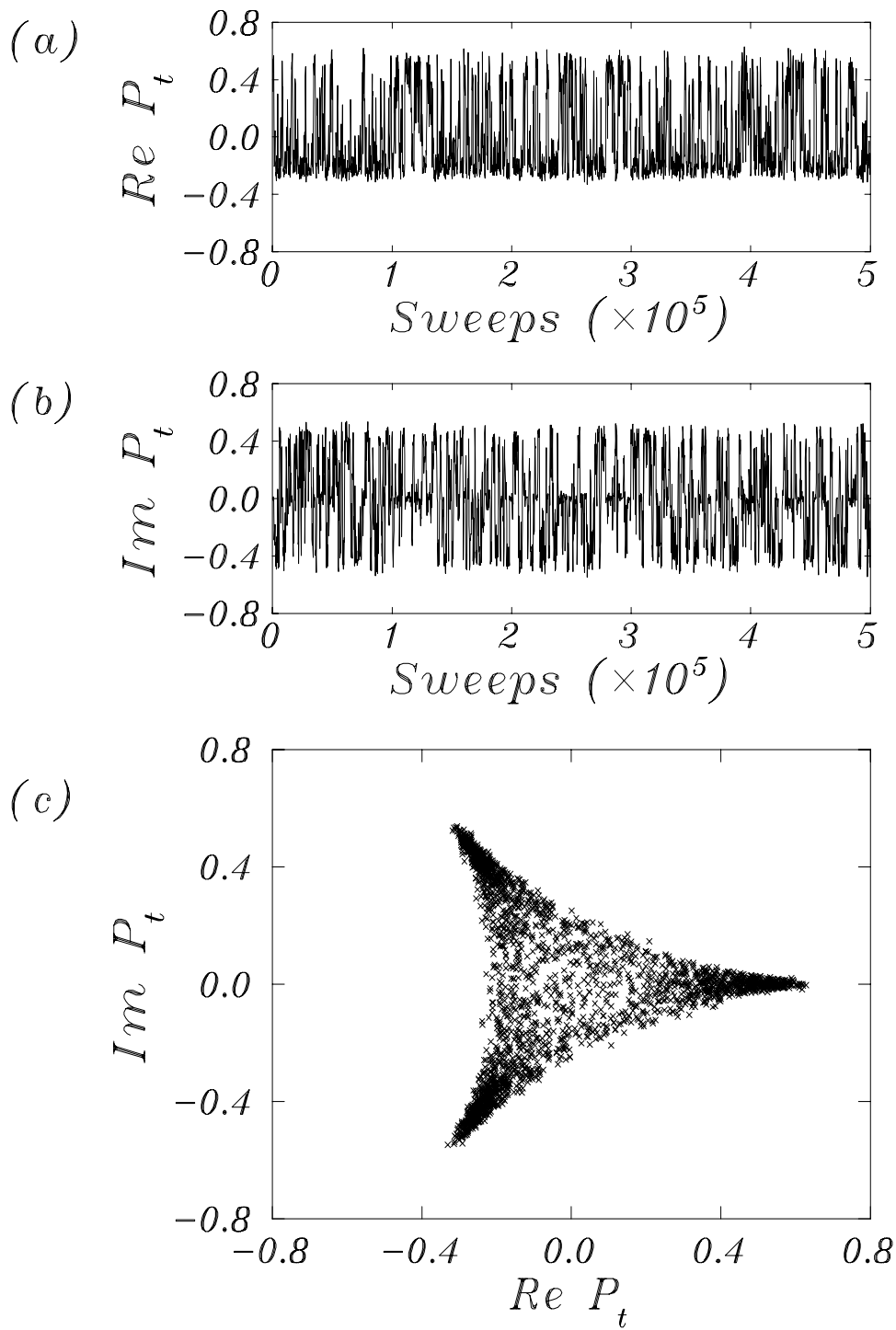,height=7.5in,width=5.in}
\vskip 0.25 in
\caption{Simulation results for the temporal Polyakov line on a $4^4$
lattice at $\beta=9$ with periodic boundary conditions. Run-time
histories are shown for (a) $\mbox{Re}\,P_t$ and (b) $\mbox{Im}\,P_t$.
A scatter plot of $\mbox{Im} \, P_t$ versus $\mbox{Re} \, P_t$
is shown in (c).}
\label{fig:TunnelPBC}
\end{figure}

\begin{figure}[htb]
\psfig{file=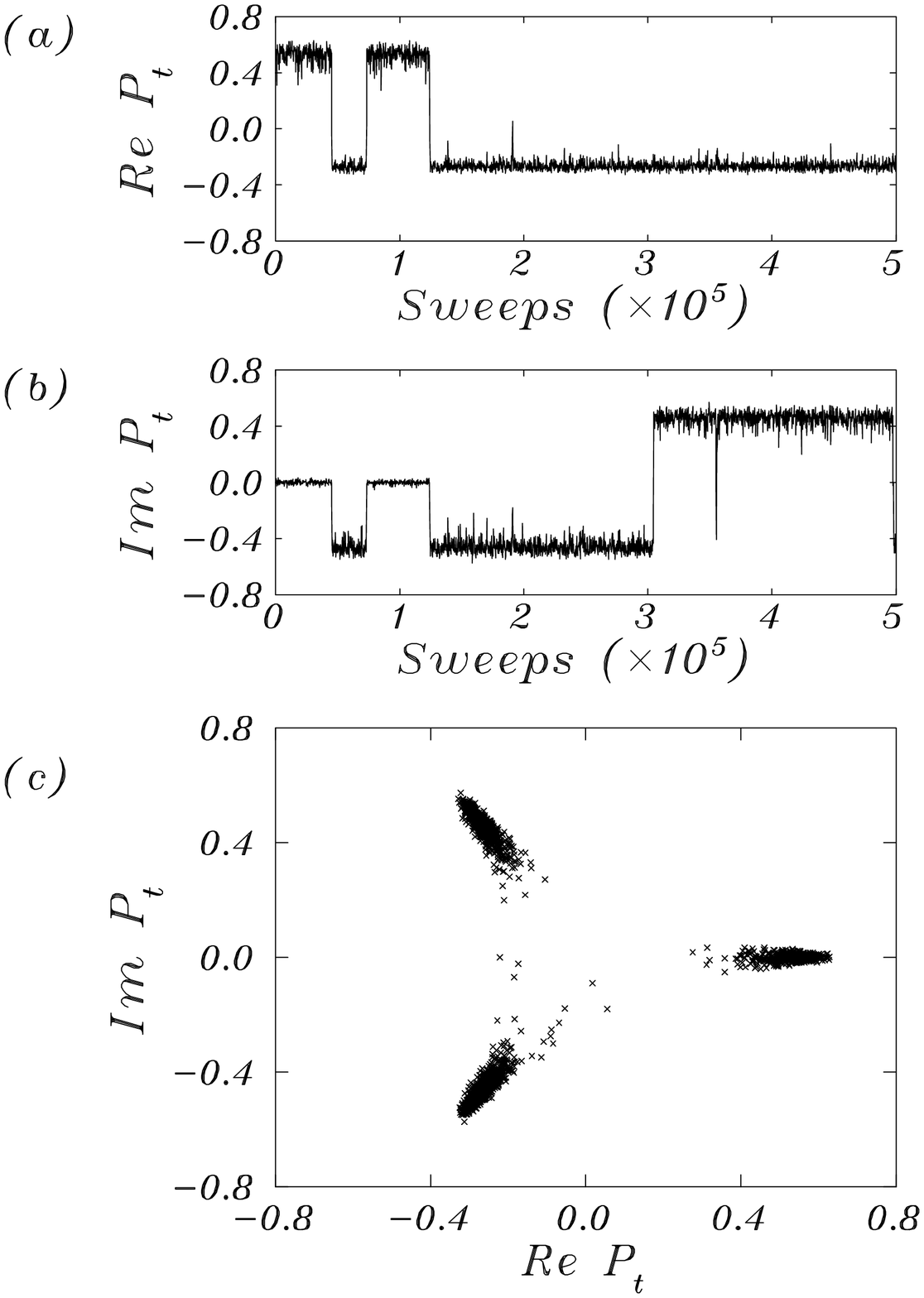,height=7.5in,width=5.in}
\vskip 0.25 in
\caption{Simulation results for the temporal Polyakov line
on a $4^4$ lattice at $\beta=9$ with twisted T$xy$ boundary
conditions. The panels are the same as in
Fig.\ \ref{fig:TunnelPBC}.}
\label{fig:TunnelTxy}
\end{figure}

\begin{figure}[htb]
\psfig{file=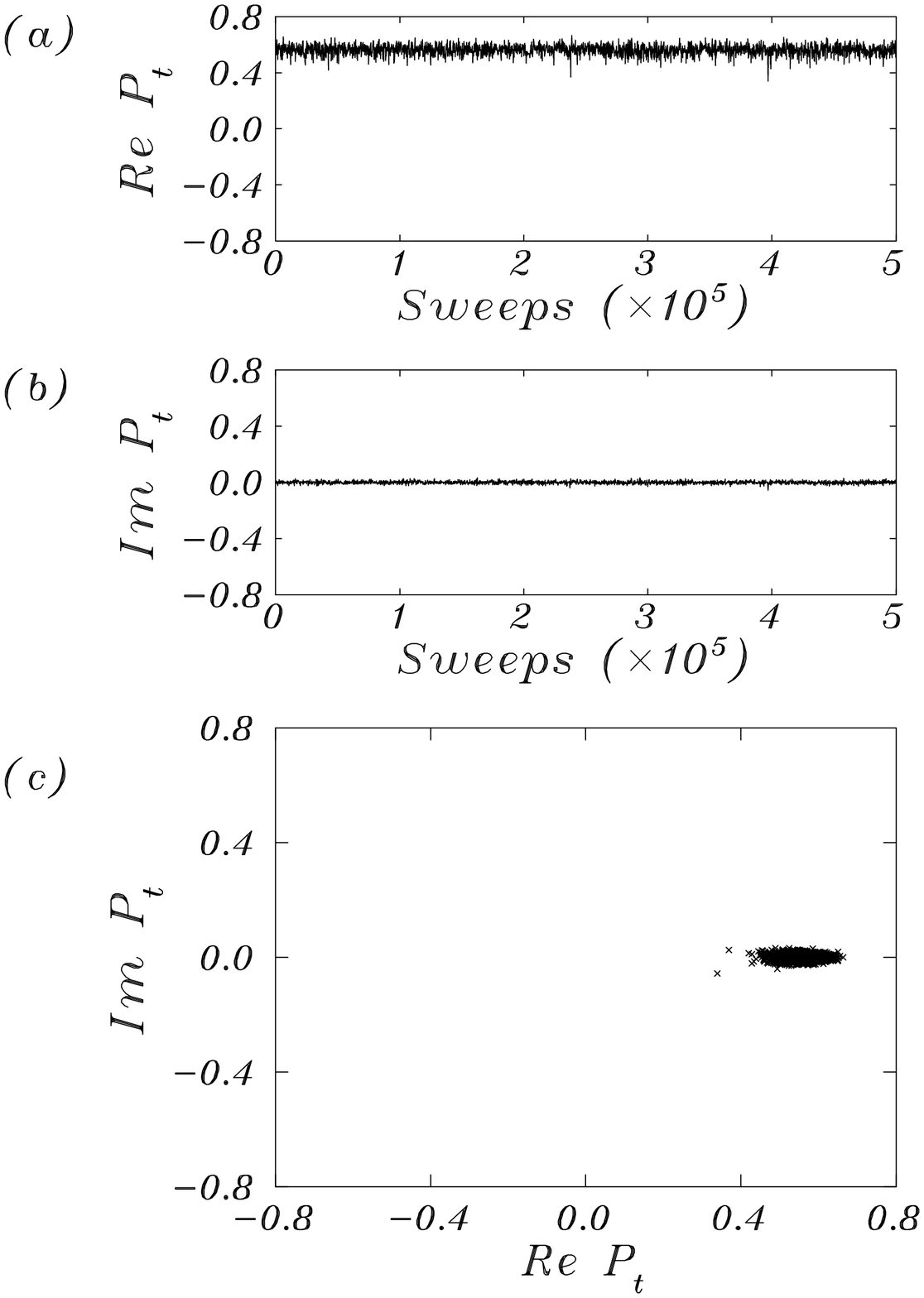,height=7.5in,width=5.in}
\vskip 0.25 in
\caption{Simulation results for the temporal Polyakov line
on a $4^4$ lattice at $\beta=9$ with twisted T$xyz$ boundary
conditions. The panels are the same as in
Fig.\ \ref{fig:TunnelPBC}.}
\label{fig:TunnelTxyz}
\end{figure}

\begin{figure}[htb]
\psfig{file=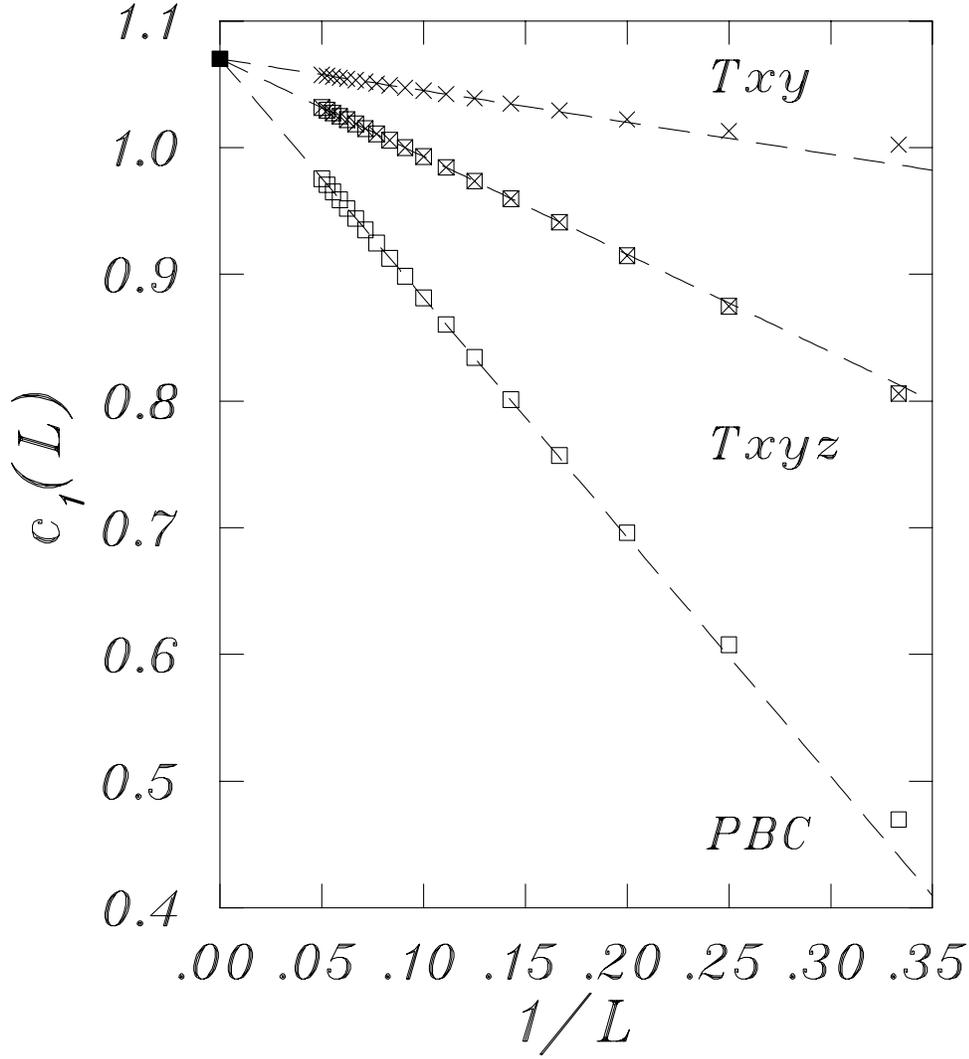,height=5.5in,width=5.in}
\vskip 0.25 in
\caption{First-order coefficient for the tadpole-improved self-energy
from perturbation theory, using different boundary conditions.
The dashed lines show fits to Eq.\ (\ref{c1Lfitform}).
The filled square shows the infinite-volume value $c_1 = 1.0701$.}
\label{fig:c1LPT}
\end{figure}

\begin{figure}[htb]
\psfig{file=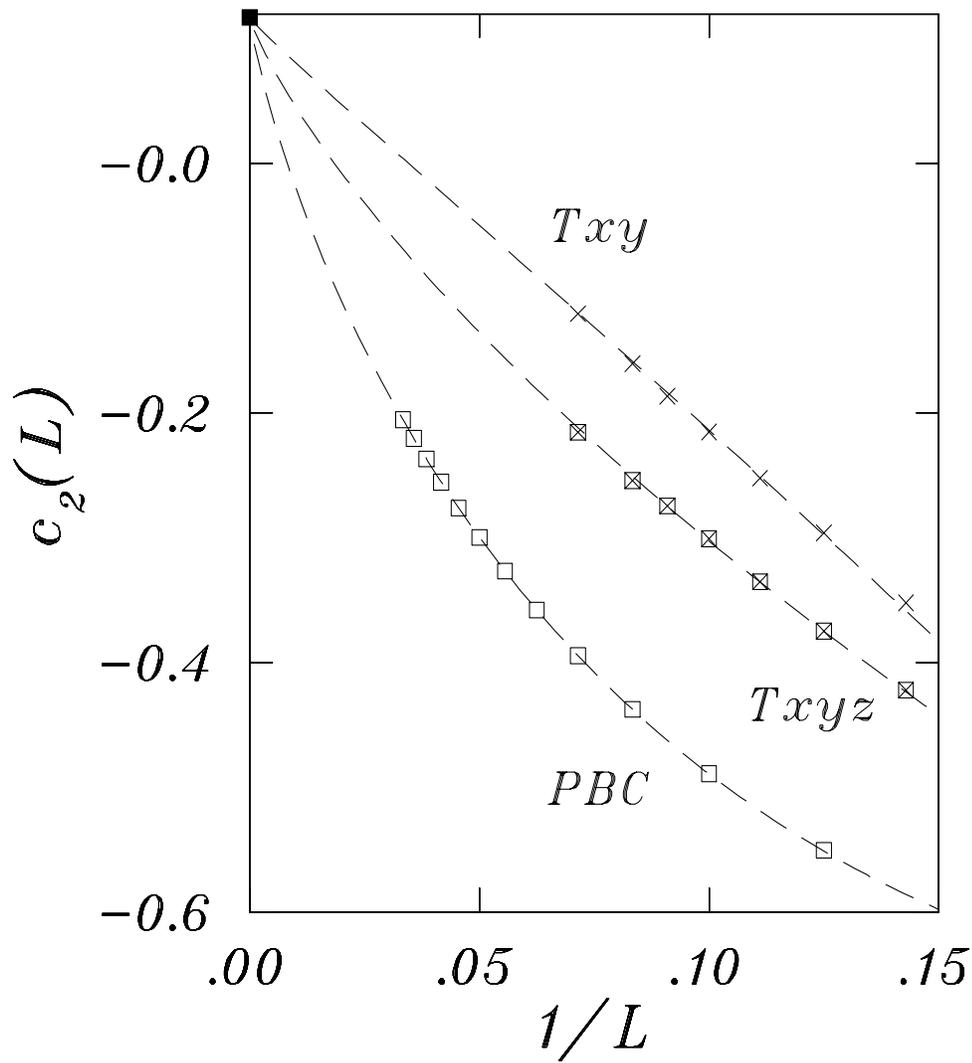,height=5.5in,width=5.in}
\vskip 0.25 in
\caption{Second-order coefficient for the self-energy from
perturbation theory, using different boundary conditions.
The dashed lines show fits to Eq.\ (\ref{c2Lfit}).
The filled square shows the infinite-volume value $c_2 = 0.117$.}
\label{fig:c2LPT}
\end{figure}

\begin{figure}[htb]
\psfig{file=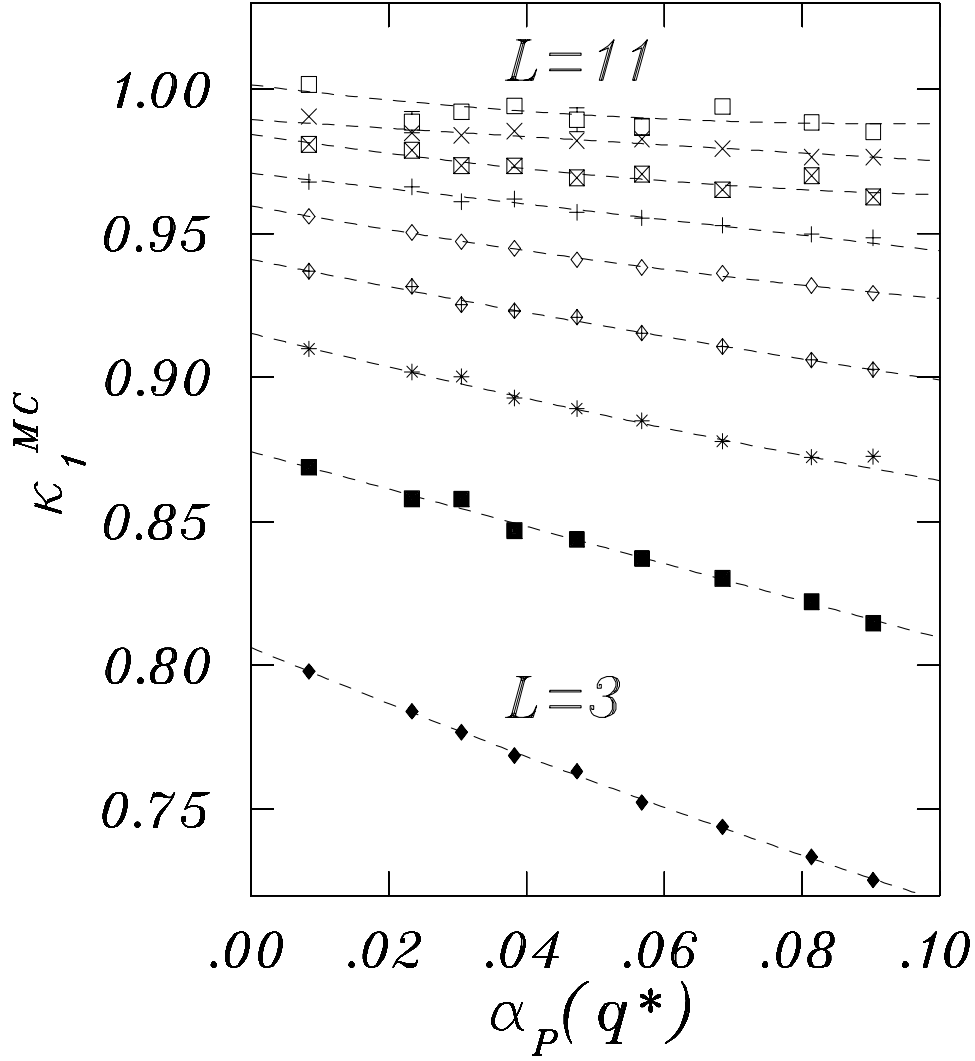,height=5.5in,width=5.in}
\vskip 0.25 in
\caption{Monte Carlo results for $\kappa_1$ for the self-energy,
Eq.\ (\ref{StaticK1}). The results for each lattice size $L$
are plotted versus the renormalized coupling $\alpha_P(q^*=0.84/a)$.
The lowest set of data points is for $L=3$, and the highest set is for
$L=11$. The dashed lines show the results of fits to
Eq.\ (\ref{E0Lseries}).}
\label{fig:StaticK1L}
\end{figure}

\begin{figure}[htb]
\psfig{file=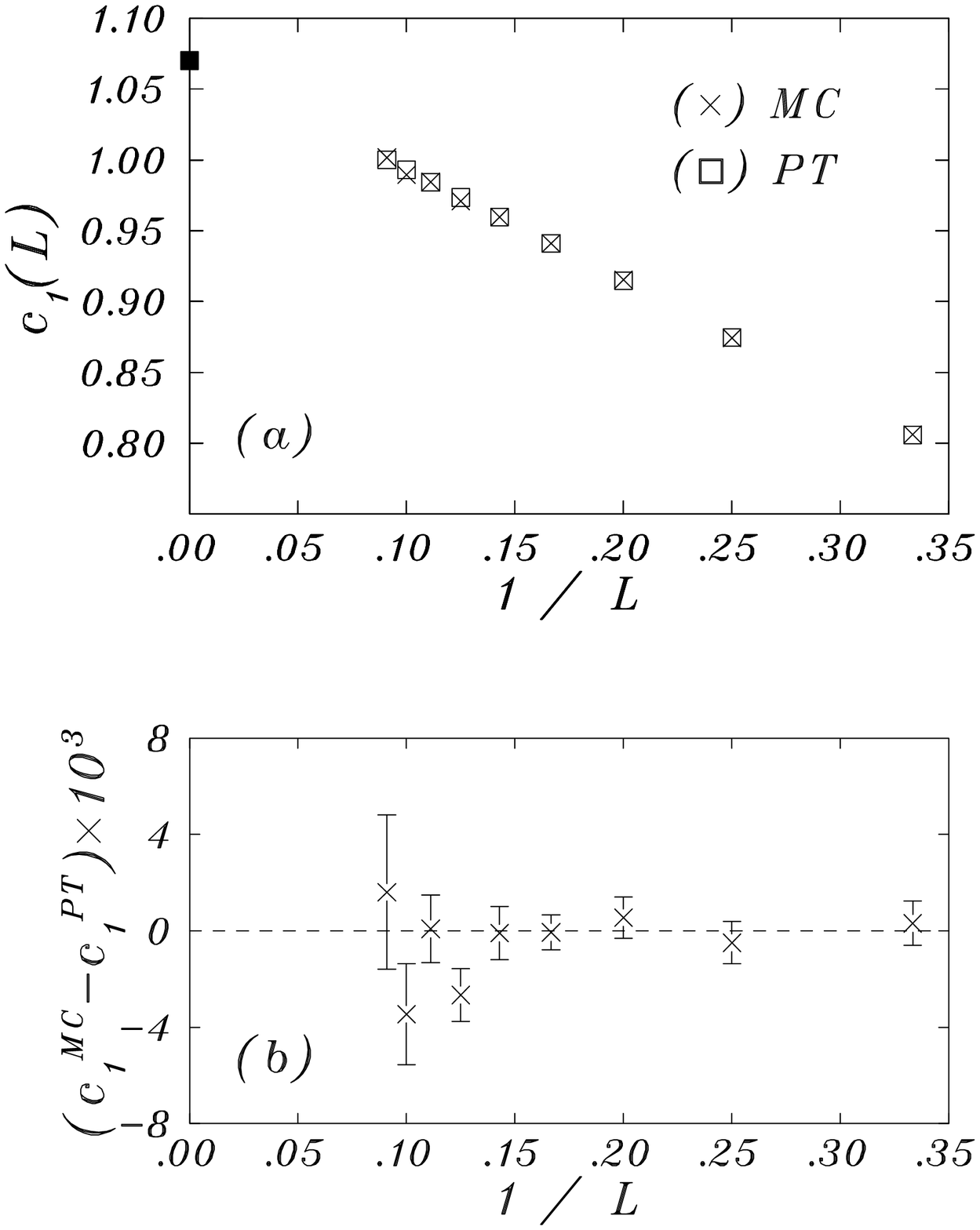,height=7.5in,width=5.in}
\vskip 0.25 in
\caption{First-order coefficient for the self-energy from Monte Carlo
simulations ($c_1^{\rm MC}$) and analytic perturbation theory
($c_1^{\rm PT}$). The filled square in (a) shows the perturbation theory
value of $c_1$ on an infinite lattice. The difference between the
Monte Carlo results and the perturbation theory is shown in (b).}
\label{fig:c1LMC}
\end{figure}

\begin{figure}[htb]
\psfig{file=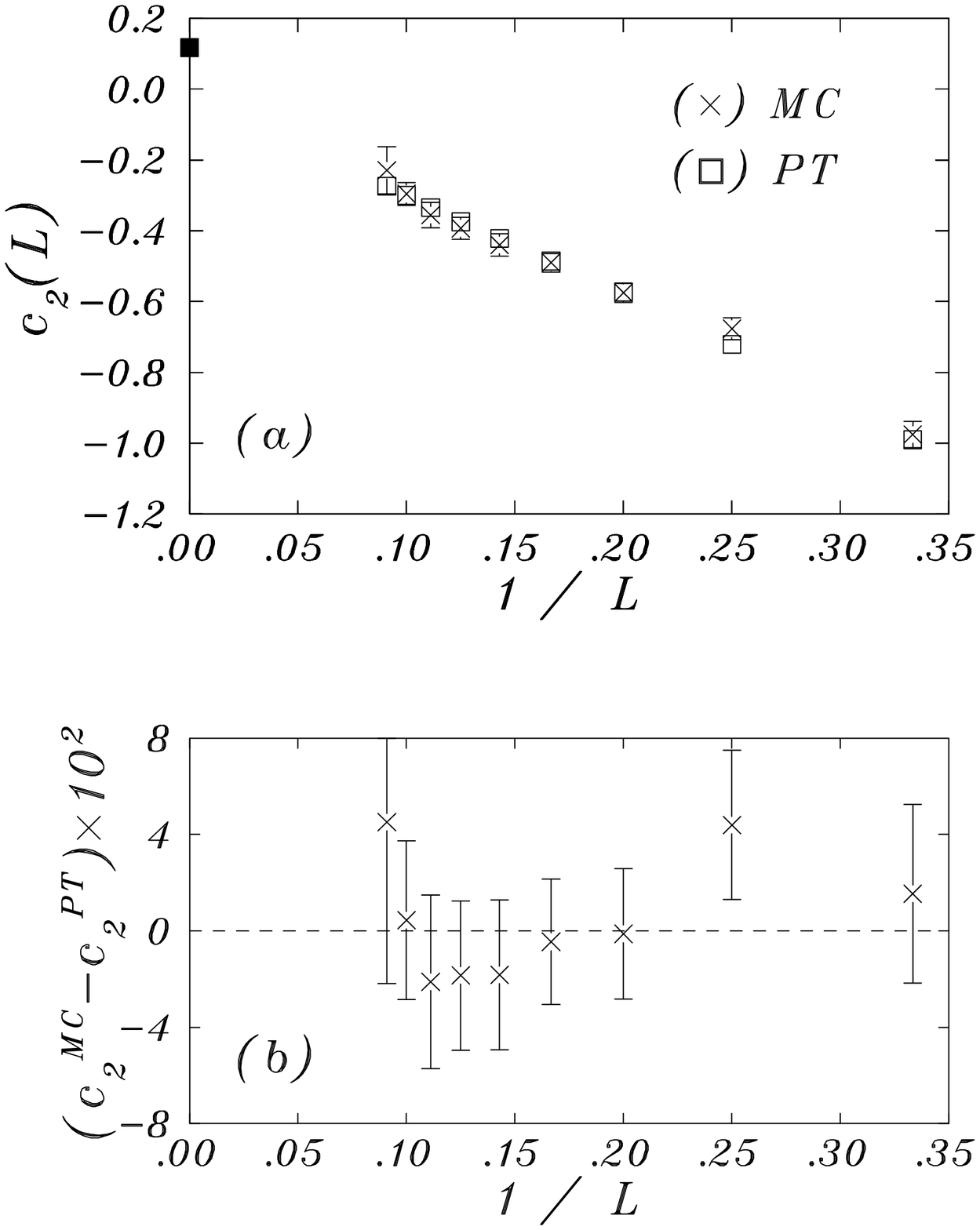,height=7.5in,width=5.in}
\vskip 0.25 in
\caption{Second-order coefficient for the self-energy from Monte Carlo
simulations ($c_2^{\rm MC}$) and analytic perturbation theory
($c_2^{\rm PT}$). The filled square in (a) shows the perturbation theory
value of $c_2$ on an infinite lattice. The difference between the
Monte Carlo results and the perturbation theory is shown in (b).}
\label{fig:c2LMC}
\end{figure}

\begin{figure}[htb]
\psfig{file=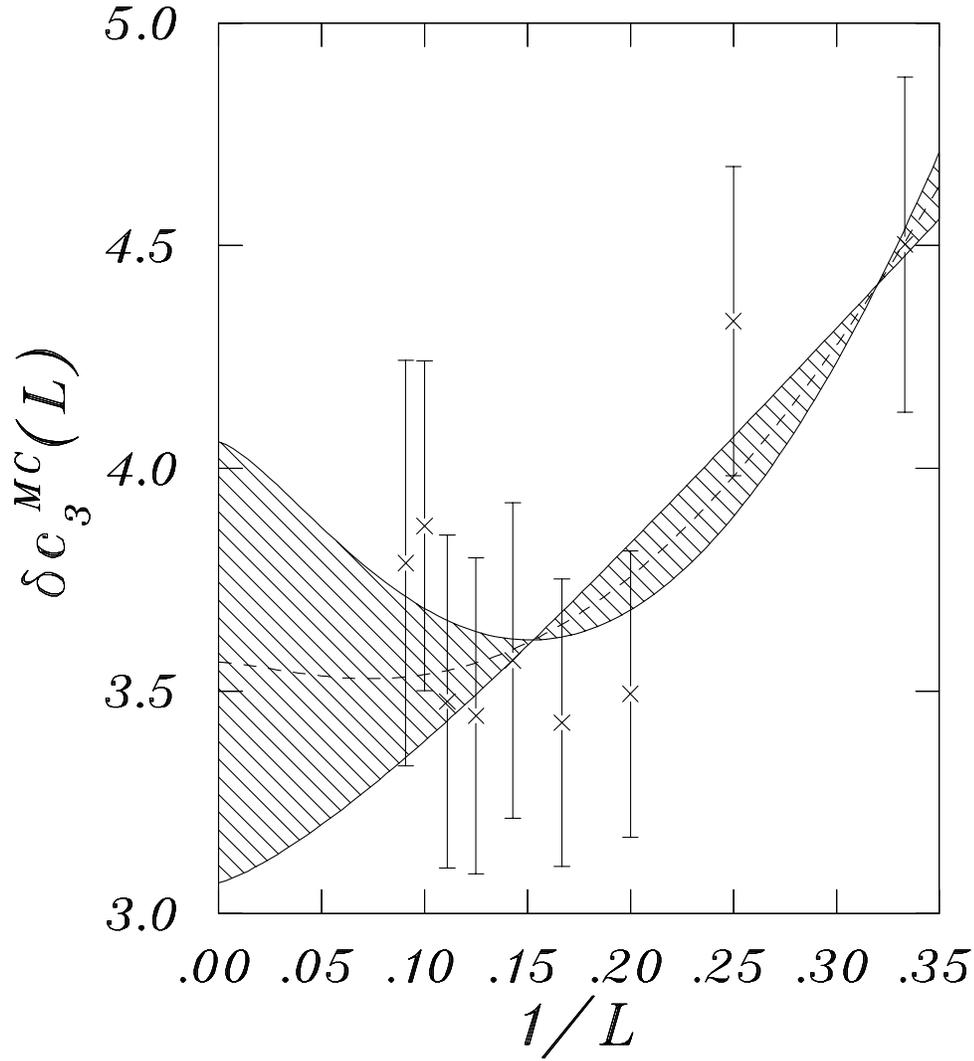,height=5.5in,width=5.in}
\vskip 0.25 in
\caption{Simulation results for the third-order coefficient for the
static-quark self-energy, after subtracting logarithms at $O(1/L)$,
according to Eq.\ (\ref{deltac3}).
The dashed line shows the result of a fit to Eq.\ (\ref{c3Lfit}),
with the shaded area corresponding to the 68\% confidence level
region for the infinite-volume coefficient.}
\label{fig:c3LMC}
\end{figure}

\begin{figure}[htb]
\psfig{file=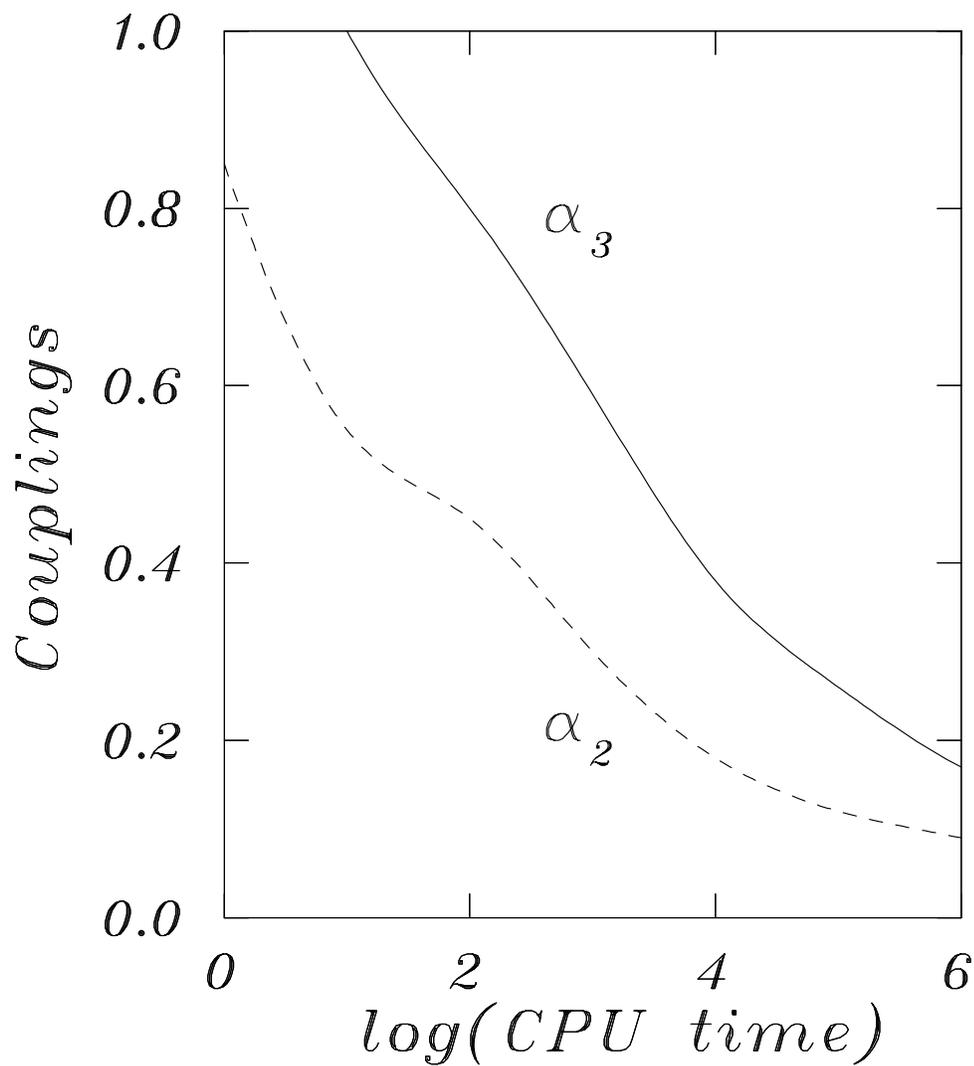,height=5.5in,width=5.in}
\vskip 0.25 in
\caption{Values of the couplings $\alpha_2$ (lower curve) and
$\alpha_3$ (upper curve) which minimize uncertainties in the
fit parameters $c_1$, $c_2$ and $c_3$, as functions of the CPU time
(in arbitrary units). The coupling $\alpha_1$ is always about zero.}
\label{fig:Aalphas}
\end{figure}

\begin{figure}[htb]
\psfig{file=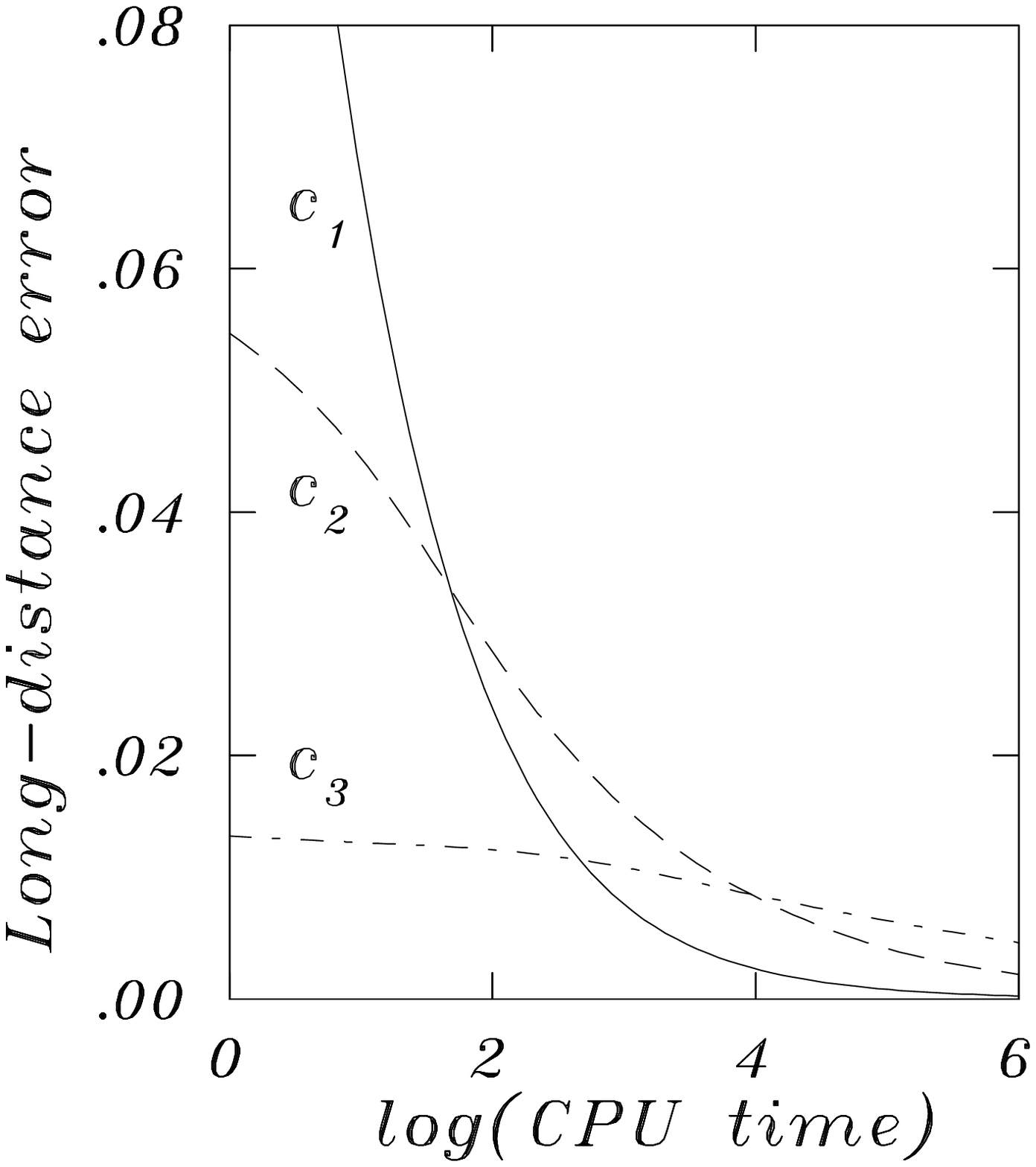,height=5.5in,width=5.in}
\vskip 0.25 in
\caption{Contribution of errors in the perturbative coefficients
to the final error in a long-distance simulation of a Wilson
loop, as a function of the short-distance simulation time
(in arbitrary units) that determined the $c_n$. The contribution
of a given coefficient $c_n$ to the final error is
$\delta c_n \alpha^n$, where here we take $\alpha=0.25$.
The errors coming from the lowest-order
coefficients are greatest for low statistics in the
short-distance simulations, but decrease most rapidly
with CPU time.}
\label{fig:Aerrors}
\end{figure}


\begin{thebibliography}{99}

\bibitem{DimmLat}
W.\ B.\ Dimm, G.\ P.\ Lepage and P.\ B.\ Mackenzie,
Nucl.\ Phys.\ B (Proc. Suppl.) {\bf 42}, 403 (1995).

\bibitem{DimmThesis}
W.\ B.\ Dimm, Ph.D. thesis, Cornell University (1995).

\bibitem{DimmWilsonLoops}
W.\ B.\ Dimm, G.\ Hockney, G.\ P.\ Lepage and P.\ B.\ Mackenzie,
unpublished [reported in
C.\ T.\ H.\ Davies et al., Phys.\ Lett.\ B {\bf 345}, 42 (1995);
Phys.\ Rev.\ D {\bf 56}, 2755 (1997).].

\bibitem{TrottierLepage}
H.\ D.\ Trottier and G.\ P.\ Lepage,
Nucl.\ Phys.\ B (Proc. Suppl.) {\bf 63}, 865 (1998).

\bibitem{Norm}
N.\ Shakespeare, Ph.D. thesis, Simon Fraser University (2000).

\bibitem{Juge}
J.\ Juge, Nucl.\ Phys.\ (Proc.\ Suppl.) {\bf 94}, 584 (2001).

\bibitem{LepMac}
G.\ P.\ Lepage and P.\ B.\ Mackenzie, Phys.\ Rev.\ D {\bf 48},
2250 (1993).

\bibitem{NRQCDalphas}
C.\ T.\ H.\ Davies et al., Ref.\ \cite{DimmWilsonLoops}.

\bibitem{PeterBayes}
G.\ P.\ Lepage, B.\ Clark, C.\ T.\ H.\ Davies, K.\ Hornbostel,
P.\ B.\ Mackenzie, C.\ Morningstar, and H.\ D.\ Trottier,
\texttt{hep-lat/0110175}.

\bibitem{tHooft}
G.\ 'tHooft, Nucl.\ Phys.\ B {\bf 153}, 141 (1979).

\bibitem{Luscher}
M.\ L\"uscher and P.\ Weisz, Nucl.\ Phys.\ B {\bf 266},
309 (1986).

\bibitem{Arroyo}
A.\ Gonzalez Arroyo and C.\ P.\ Korthals Altes,
Nucl.\ Phys.\ B {\bf 311}, 433 (1988).

\bibitem{Lat99}
Some of this work was presented in preliminary form in G.\ P.\ Lepage,
P.\ B.\ Mackenzie, N.\ H.\ Shakespeare and H.\ D.\ Trottier,
Nucl.\ Phys.\ (Proc.\ Suppl.) {\bf 83}, 866 (2000);
and in Ref.\ \cite{Norm}.

\bibitem{Heller}
U.\ M.\ Heller and F.\ Karsch, Nucl.\ Phys.\ B251 (1985) 254.

\bibitem{Martinelli}
See also G.\ Martinelli and C.\ Sachrajda,
Nucl.\ Phys.\ B {\bf 559}, 429 (1999).

\bibitem{OurTwistedPT}
H.\ D.\ Trottier, G.\ P.\ Lepage, Q.\ Mason, M.\ A.\ Nobes,
and K.\ Foley, in preparation. See also
M.\ A.\ Nobes, H.\ D.\ Trottier, G.\ P.\ Lepage and Q.\ Mason,
\texttt{hep-lat/0110051}.

\bibitem{E0mbSims}
See e.g. V.\ Gimenez, L.\ Giusti, G.\ Martinelli and
F.\ Rapuano, JHEP {\bf 0003}, 018 (2000);
S. Collins, hep-lat/0009040; and
Ref.\ \cite{Martinelli}.

\bibitem{Coste}
A.\ Coste et al., Nucl.\ Phys.\ B {\bf 262}, 67 (1985).

\bibitem{ParmaE0}
G.\ Burgio, F.\ Di~Renzo, M.\ Pepe, and L.\ Scorzato
Nucl.\ Phys.\ (Proc.\ Suppl.) {\bf 83}, 935 (2000);
F.\ Di~Renzo and L.\ Scorzato, JHEP {\bf 0102}, 020 (2001).

\bibitem{Alles}
B.\ All\`es et al., Phys.\ Lett.\ B {\bf 324}, 433 (1994).

\bibitem{PDG} See e.g. I. Hinchliffe in
Particle Data Group, D.\ E.\ Groom {\it et al.},
Eur.\ Phys.\ J.\ C {\bf 15}, 1 (2000).

\bibitem{LuscheralphaMS}
M.\ L\"uscher and P.\ Weisz, Phys.\ Lett.\ B {\bf 349}, 165 (1995);
Nucl.\ Phys.\ B {\bf 452}, 234 (1995).

\bibitem{Rodrigo}
For this purpose we use a solution to the evolution equation
that explicitly gives the dependence of the running coupling
on the coupling at the reference scale.
See, for example, G.\ Rodrigo, A. Pich,
and A. Santamaria, Phys.\ Lett.\ B {\bf 424}, 367 (1998); and
G. Rodrigo and A. Santamaria, Phys.\ Lett.\ B {\bf 313}, 441 (1993).

\bibitem{ParmaWilson}
F.\ Di~Renzo, E.\ Onofri and G.\ Marchesini,
Nucl.\ Phys.\ B {\bf 457}, 202 (1995).


\end{thebibliography}
\end{document}